\documentstyle[12pt]{JHEP3}
\newcommand{\be}[1]{ \begin{equation}\label{#1} }
\newcommand{\ee}{\end{equation}}
\newcommand{\bea}[1]{\begin{eqnarray}\label{#1} }
\newcommand{\eea}{\end{eqnarray}}
\newcommand{\eq}[1]{(\ref{#1})}
\def\ZZZ{{\hskip-3pt\hbox{ Z\kern-1.6mm Z}}}
\def\zzz{{\hskip-3pt\hbox{ z\kern-1mm z}}}

\newcommand{\gl}{\lambda}

\newcommand{\gp}{\psi}
\newcommand{\gt}{\theta}

\newcommand{\gk}{\chi}

\newcommand{\GG}{\Gamma}
\newcommand{\hf}{{1\over 2}}
\newcommand{\thf}{{3\over 2}}

\newcommand{\half}{{1\over 2}}

\newcommand{\sldel}{\slash{\!\!\!\! \nabla}}

\def\one{{\hbox{ 1\kern-.8mm l}}}
\def\zero{{\hbox{ 0\kern-1.5mm 0}}}

\def\Tr{\,{\rm Tr}\, }

\def\hn{\hat{n}}
\def\tg{\tilde{g}}

\title{
The Heat Kernel on AdS$_3$ and its Applications}

\author{
Justin R.\ David$^a$\footnote{On leave 
from Harish-Chandra Research Institute, Allahabad.},
Matthias R.\ Gaberdiel$^b$ and Rajesh Gopakumar$^c$ \\ 
$^a$Centre for High Energy Physics,\\
$\;$Indian Institute of Science, \\
$\;$Bangalore 560012, India\\
$\;$\email{justin@cts.iisc.ernet.in}\\ \\
$^b$Institut f\"ur Theoretische Physik, ETH Zurich, \\
$\;$CH-8093 Z\"urich, Switzerland \\
$\;$\email{gaberdiel@itp.phys.ethz.ch}\\ \\ 
$^c$Harish-Chandra Research Institute, \\
$\;$Chhatnag Road,\\
$\;$Jhusi, India 211019\\
$\;$\email{gopakumr@hri.res.in}}

\abstract{
We derive the heat kernel for arbitrary tensor fields  on $S^3$ and (Euclidean) AdS$_3$ using a 
group theoretic approach. We use these results to also obtain the heat kernel on certain quotients 
of these spaces. In particular, we give a simple, explicit expression for the one loop determinant  
for a field of arbitrary spin $s$ in thermal AdS$_3$.  We apply this to the calculation of the 
one loop partition function of ${\cal N}=1$ supergravity on AdS$_3$. We find that the answer 
factorizes into left- and right-moving super Virasoro characters 
built on the $SL(2, {\mathbb C})$ invariant vacuum, as argued by Maloney and Witten on 
general grounds. }

\preprint{HRI/ST/0922}

\begin{document}

\section{Introduction}

In studying the quantization of field theories on a general spacetime an important tool which 
captures the leading quantum properties of the theory is the heat kernel of the Laplacian. 
Even if the full quantum theory is ill-defined or ill-understood (as is the case for theories of gravity), 
this leading one loop behaviour is
typically well defined and often under analytic control. Knowing the heat kernel enables one to 
compute, for instance, the one loop determinants that contribute to the free energy.  The heat 
kernel also contains the information about the propagator and other important one 
loop effects such as the  anomalies of the quantum theory. 

In these notes we will study the heat kernel on (Euclidean) AdS$_3$ spacetime for particles of 
arbitrary spin $s$. In studying the leading quantum effects for pure gravity or supergravity on 
AdS$_3$ one needs to compute the heat kernel for particles with  spin less than or equal to two. 
More generally, for a string theory on AdS$_3$ one would need the heat kernel for particles of 
arbitrary spin $s$. With a view to some of these potential applications we obtain  
expressions for the heat kernel  of the Laplacian $\Delta_{(s)}$ acting on tensor fields 
(transverse and traceless of arbitrary spin $s$). 
We will give answers for the cases of $S^3$ and some simple quotients as well 
as for Euclidean AdS$_3$ ({\it i.e.}\ $H_3^+$) and its thermal quotient. In particular, we obtain 
explicit expressions for the heat kernel for coincident points whose integral over proper time  
gives the one loop determinant. 

As an immediate application of these results we are able to evaluate the one loop 
contribution from the physical spin $\thf$ gravitino in, for example, 
${\cal N}=1$ supergravity on thermal AdS$_3$. This one loop result together with the 
answer for the spin two graviton combines into left- and right-moving super-Virasoro 
characters for the identity representation
\be{thf1loop}
Z_{{\rm 1-loop} } =  
\prod_{n=2}^\infty \frac{ | 1 + q^{n - \frac{1}{2}} |^2  }{|1- q^n|^2} \ ,
\ee
where $q=e^{i\tau}$ parametrizes the boundary $T^2$ of the thermal AdS$_3$. 
This agrees with the general argument given by Maloney and Witten \cite{Maloney:2007ud}
which was based on an extension of the results of Brown and Henneaux 
\cite{Brown:1986nw}. Maloney and Witten  in fact also argued that (in an appropriate 
choice of scheme) this result was perturbatively one loop exact. The bosonic version of this 
argument for pure gravity (the denominator term in \eq{thf1loop}) has been checked by the 
computation of Giombi {\it et.al.}\ \cite{Giombi:2008vd} who have explicitly evaluated the 
heat kernel for transverse vectors and spin two fields. Our results for the supergravity case 
complete this check of the Maloney-Witten argument. 
\smallskip

We now give a broad overview of our methods. As mentioned above, 
the heat kernel for AdS$_3$ and its thermal quotient have been explicitly evaluated
for transverse vectors and spin two tensors \cite{Giombi:2008vd}. 
The method of evaluation employed there is however fairly cumbersome to 
generalise to arbitrary spin. We will instead adopt a more geometric 
approach. We will exploit  the fact that $S^3=SU(2)=(SU(2)\times SU(2))/SU(2)$ 
and $H_3^+=SL(2,{\mathbb C})/SU(2)$ are homogeneous spaces. The fields of arbitrary spin 
$s$ are therefore sections of what are known as homogeneous vector bundles on these 
coset spaces. This will allow us to use some well-known techniques of harmonic analysis 
to write down the eigenfunctions of the spin $s$ Laplacian $\Delta_{(s)}$  in terms of matrix elements of 
representations of $SU(2)\times SU(2)$ and $SL(2,{\mathbb C})$. These have, in fact, 
already appeared in the physics literature in a series of papers by Camporesi and Higuchi
\cite{Higuchi:1986wu,Camporesi:1990wm,Camporesi:1992tm,Camporesi:JGP,%
Camporesi:1994ga,Camporesi:1995fb} (see also 
\cite{Dowker66,Dowker83,Chodos:1983zi,Ordonez1984, Vassil} for some related work).
We will heavily draw upon these methods and adapt them to obtain the expressions 
of  interest to us. 

Given the eigenfunctions of the Laplace operator we can evaluate the heat kernel as 
\be{heatkdef}
K^{(s)}_{ab}(x,y; t) = 
\langle y,b |e^{t\Delta_{(s)}} |x, a\rangle 
=\sum_{n}\psi^{(s)}_{n,a}(x)\, \psi^{(s)}_{n,b}(y)^*\, e^{t\lambda_n^{(s)}}
\ee
for arbitrary pairs of points $(x,y)$ on the space in question ($S^3$ or $H_3^+$). 
Here $a,b$ are labels for the $2s+1$ dimensional representation for spin $s$. 
The eigenfunctions $\psi_n$ have been labelled by $n$, which will denote a 
multi-index, while $\lambda_{n}^{(s)}$ is the corresponding eigenvalue.  
Using the group theoretic origin of the wave functions $\psi^{(s)}_{n,a}(x)$ 
we can carry out partial sums over degenerate eigenstates (those having the same eigenvalue 
$\lambda_{n}^{(s)}$). This manifests itself as a generalised version of the addition theorems
that make their appearance in special function theory.

Given the heat kernel one can compute the one loop determinant, for instance, by considering the 
coincident limit of the heat kernel
\be{oneloopdet}
\ln {\rm det}(-\Delta_{(s)}) 
={\rm Tr} \ln(-\Delta_{(s)}) = -  \int_0^{\infty}{dt\over t} \int \sqrt{g}\, d^3x \, K^{(s)}_{aa}(x,x; t)\ .
\ee
To compute the heat kernel, as well as one loop determinants, on
quotients of $S^3$ or $H_3^+$ we can use the method of images. 
The basic quotients we will study are Lens space quotients of $S^3$ while the analogous 
quotient in $H_3^+$ is the one giving Euclidean thermal AdS.\footnote{In the case of $H_3^+$ and its 
thermal quotient the expression in \eq{oneloopdet} suffers from a trivial volume divergence which
we will ignore; 
we shall concentrate on the finite piece which contains all the nontrivial $q$ dependence.}

We will describe the $S^3$ case (and its quotient) in great detail in Secs.\ 2, 3 and 4, both 
because it is compact and because many of the group theoretic features use only familiar facts about 
representations of $SU(2)$.  In Sec.\ 2 we briefly summarize some of the relevant 
ideas from harmonic analysis which lead to the explicit forms of the eigenfunctions of the 
spin $s$ Laplacian. We go on to give a number of different expressions for these eigenfunctions 
as well as their explicit form for low values of the spin. Sec.\ 3 uses these expressions and 
their group theoretic origin to write down the heat kernel for separated points. Once again 
a number of explicit expressions are worked out. Sec.\ 4 deals with a Lens space like quotient 
of $S^3$ and the method of images is applied to obtain the heat kernel. 

The case of $H_3^+$ is more subtle since it involves harmonic analysis on a non-compact group. 
The relevant representations are infinite dimensional, and the discrete sums 
in \eq{heatkdef} become continuous integrals with an appropriate measure. While these 
are relatively well understood in the case of interest to us, namely $SL(2,{\mathbb C})$, we will 
practically implement the calculation by performing a suitable analytic continuation of the 
answers from $S^3$. Analytic continuation from compact to non-compact groups is often fraught 
with danger, and one needs to proceed with caution. In this case, however, it is known from 
works of Helgason \cite{Helgason} and Camporesi-Higuchi \cite{Camporesi:JGP, Camporesi:1995fb}
that analytic continuation works. In fact, $S^3$ and $H_3^+$ are among the simplest examples of 
`dual spaces' on which harmonic analysis can be analytically continued. We will elaborate on 
this in Sec.\ 5. In Sec.\ 6, we extend this analytic continuation to thermal quotients of 
$S^3$ and $H_3^+$ and obtain an explicit and relatively simple expression for the 
(integrated and coincident) heat kernel (see eq.~\eq{thermalh3-ker}). We check that this answer 
correctly reproduces all the previously known cases ({\it i.e.}\ spins $s=0,1,2$). 

Finally, in Sec.\ 7, we use the results of Sec.\ 6 to evaluate the one loop 
partition function of ${\cal N}=1$ supergravity on AdS$_3$. This additionally 
requires a careful analysis of the physical quadratic fluctuations of the 
massless gravitino about the AdS$_3$ background. We carry this out and 
show that the final answer takes the expected form \eq{thf1loop}. Various additional
details are relegated to the four appendices.

\section{Construction of Harmonics  on $S^3$}

We will be interested in the symmetric traceless divergence free (transverse) 
 tensors of spin $s$ on $S^3$. 
This is sufficient information to study fields in arbitrary representations.\footnote{Note that since we 
are working in three dimensions there are no non-trivial antisymmetric representations that need to 
be considered: the two form is dual to a vector and the three form to a scalar.} To construct the heat 
kernel we need the complete set of eigenfunctions of the corresponding Laplacian $\Delta_{(s)}$. 
This can be explicitly studied  using  harmonic analysis on homogeneous vector bundles which 
applies directly to homogeneous spaces of the form $G/H$  (see \cite{Camporesi:1995fb} for 
an accessible introduction for physicists). The harmonic wavefunctions can be expressed in 
terms of matrix elements of particular representations of $G$.  We will start by considering the 
case where $G$ is compact as exemplified by 
$S^3$ which can be thought of as the homogeneous space
\be{S3}
S^3 \cong (SU(2)\times SU(2))/SU(2)\ ,
\ee
with the denominator acting diagonally on $(SU(2)\times SU(2))$, {\it i.e.}
\be{coset}
(g_L,g_R) \mapsto (g_L \cdot h,g_R \cdot h) \  , \qquad h \in SU(2) \ .
\ee
 We can identify the quotient space, via the projection map $\pi$,  with $SU(2)=S^3$ itself,
\be{coseta}
\pi: SU(2)\times SU(2) \rightarrow SU(2) \ , \qquad
(g_L,g_R) \mapsto g_L \cdot g_R^{-1} \ .
\ee
This map is evidently independent of the representative, {\it i.e.}\ it 
is invariant under replacing $(g_L,g_R)$ by 
$(g_L\cdot h ,g_R\cdot h)$. 
\smallskip
Below we will describe the corresponding tensor harmonics on $S^3$
in terms of matrix elements of $SU(2)\times SU(2)$. 

To write explicit expressions we will also need to choose definite coordinates on $S^3$. 
The most common set of coordinates is the spherical system parametrized by 
$(\chi, \theta,\phi)$
in which the metric of $S^3$ reads 
\be{sphercoords}
ds^2=d\chi^2+\sin^2\chi \, (d\theta^2+\sin^2\theta \, d\phi^2) \ .
\ee
The corresponding group element in $SU(2)$ is parametrized by
\be{spheref1}
g(\chi, \theta,\phi) = \left( \begin{array}{cc}
\cos{\chi} + i\sin{\chi} \cos{\theta} & 
i \sin\chi\, \sin\theta\, e^{i\phi} \\
i\sin\chi\, \sin\theta\, e^{-i\phi}  & 
\cos{\chi} - i\sin{\chi} \cos{\theta}
\end{array} \right) \ .
\ee
This will be useful for comparing some of the results to known expressions in the literature. 

However, for performing the thermal quotient it will be most convenient to use double polar coordinates
$(\psi, \eta, \varphi)$ in terms of which the metric reads 
\be{torcoords}
ds^2=d\psi^2+\cos^2\psi\, d\eta^2+\sin^2\psi \, d\varphi^2\ .
\ee
In terms of these coordinates the elements of $SU(2)$ are given by
\be{torusf}
g(\psi, \eta, \varphi) = \left(
\begin{array}{cc}
e^{-i\eta} \cos\psi \quad & i e^{i\varphi} \sin\psi \\
i e^{-i\varphi} \sin\psi  \quad & e^{i\eta} \cos\psi
\end{array}
\right) \ .
\ee

\subsection{Tensor Harmonics and Representation Theory}

The nature of $S^3$ as a homogeneous space allows one to choose tensor harmonics 
with respect to a basis which reflects this homogeneity (see below). Though focussing on 
$S^3$ (and later $H_3^+$)  many of the ideas are general and we will often indicate the 
generalization to general homogeneous spaces.
We refer to  \cite{Camporesi:1995fb} for a more comprehensive discussion.

An important role will be played by sections 
$\sigma(x)$  of the principal bundle $SU(2)\times SU(2)$ over the base $SU(2)$ 
(being parametrized by $x$). That is
\be{sigmad}
\sigma: SU(2) \rightarrow SU(2)\times SU(2) \ , \qquad
\hbox{such that} \qquad
\pi \circ \sigma = {\rm id}_{SU(2)} \ .
\ee
Obviously, there is no canonical choice of a section. In particular, for any given
$\sigma$, we can define $\hat\sigma$ via 
\be{hsig}
\hat\sigma = \sigma \cdot (h(x),h(x)) \ , 
\ee
where $h(x)$ is any map from $SU(2)\rightarrow SU(2)$. From the definition of 
the quotient action (\ref{coset}), it is clear that any two sections are related in this 
manner. 

Any given section $\sigma(x)$ actually also determines a natural choice for a basis of 
tensor valued functions.  Define ${\bf v}_a$  $(a=1 \ldots 2s+1)$ 
as a basis for a spin $s$ 
representation of $SU(2)$ at the origin (of $S^3$ viewed as a group).
Then a basis of sections of the spin $s$ tensor bundle can be defined via
\be{sigmaten}
\theta_a(x) = \sigma(x) {\bf v}_{a} \ .
\ee
For the case of spin $s=1$, ${\bf v}_a$ can be thought of as a vector in the tangent 
space of $SU(2)\times SU(2)$ at the identity, and the action of 
$\sigma(x)\in SU(2)\times SU(2)$  is the usual push-forward. 
The form of the resulting vielbein basis, for some of the sections that we will use, is 
summarized in Appendix B. The generalization to arbitrary spin $s$ is then straightforward.

We will expand our tensor harmonics in this basis. 
\begin{equation}
\Psi(x)=\sum_a\Psi_a(x)\theta_a(x)\ .
\end{equation}
In other words, it is the components
$\Psi_{a}(x)$ (with respect to the basis $\theta_a(x)$) which will be 
the eigenfunctions of the Laplace operator $\Delta_{(s)}$.  
The arbitrariness we saw above in the choice of the section reflects 
a freedom in the choice of basis (see Appendix A for more details).
We will see below that this freedom will be reduced in the presence of quotients. 
The tensor harmonics that will be explicitly given below are always defined
with respect to some basis $\{\theta_\alpha (x) \}$ determined by a 
particular choice of section. 

Having identified the basis of tensors, we can now give explicit
formulae for the component tensor harmonics  \cite{Camporesi:1995fb}. 
Here we will describe the approach for a general compact homogeneous space. 
Geometrically, the tensors we are considering are sections of 
homogeneous vector bundles $E_{\rho}$ associated to the principal bundle $G$
over the homogeneous space $G/H$, with structure group $H$ and transforming under 
some particular representation $\rho$  of $H$. The harmonic analysis of such vector bundles 
is an extension of the usual harmonic analysis for scalars. 

The crucial point we shall use is that there is a natural
embedding of the space of sections of these bundles into the space of functions on $G$.
We can make this correspondence one to one if we restrict ourselves to 
the functions $\psi_a(g)$ on $G$,\footnote{Technically, this is 
the statement that $L^2(G)$ decomposes into a union (over representations $\rho$, 
with some multiplicity) of the spaces $L^2(G/H, E_{\rho})$. This is familiar to physicists 
in the study of monopole harmonics on $S^2$ ($G=SU(2), H=U(1)$) all of which 
arise from (equivariant) functions on $S^3$.} that are {\it equivariant}
with respect to $H$. These functions obey
\be{equivariant}
\psi_a(gh)=\rho(h^{-1})_a^b\, \psi_b (g)
\ee
for any $g\in G$ and $h\in H$, where $\rho(h)$ is the representation of $H$ acting 
on the fibres of the vector bundle. We can thus think of the $\psi_a(g)$ as components 
of a vector which lie in the 
vector space of a typical fibre ({\it e.g.}\ at the origin with respect to a basis 
$\{ {\bf v}_a \}$ in our case) of the associated vector bundle.

Now we can use the section $\sigma(x)$ of the principal fibre bundle $G$
to construct tensor valued component functions on $G/H$ (with respect to the basis $\theta_a(x)$
arising from the section $\sigma(x)$ as in (\ref{sigmaten}))
via
\be{secfunc}
\Psi_a(x)=\psi_a(\sigma(x))\ .
\ee
In our case, with $g\in G=SU(2)\times SU(2)$, is not difficult to see that the functions 
\be{genhar}
\psi^{(\lambda; I)}_a(g) = U^{\gl}(g^{-1})^I_a \ ,
\ee
are equivariant with respect to $H=SU(2)$.
Here  $\gl$ denotes a 
representation of $SU(2)_L\times SU(2)_R$ 
which contains the spin $s$ representation under the diagonal action of $SU(2)$. 
The label $a$ takes values in the spin $s$ representation
that is contained in $\lambda$ under the diagonal action, while $I$ labels
the different states in the representation $\lambda$. Finally,
$U^\lambda$ denotes the matrix elements of the unitary representation
$\lambda$. We shall exhibit this formula more explicitly below, see
(\ref{psidef}) and (\ref{psimdef}). There is 
also an obvious generalization of this for arbitrary 
$G$ and $H$. 

For each such choice of $\lambda$, we can thus write down, using the above correspondence
\eq{secfunc}, the 
components of a tensor section as 
\be{genhar1}
\Psi^{(\lambda; I)}_a(x) = U^{\gl}(\sigma(x)^{-1})^I_a \ .
\ee
In fact, these components of (\ref{genhar}) 
are actually eigenfunctions of the spin $s$ Laplacian 
(with the conventional spin connection in the covariant derivative)  for each state 
in $\lambda$ (labelled by $I$) \cite{Camporesi:1995fb}. 
These constitute a complete set of rank $s$ tensor harmonics, whose components 
(with respect to the basis (\ref{sigmaten})) are described by the index $a$. 
In order to describe the {\em transverse and traceless} tensors of spin $s$ the representations 
$\lambda$ must be taken to be of the form \cite{Ordonez1984,Higuchi:1986wu}
\be{reps}
\lambda_+ = \left(\frac{n}{2}+s, \frac{n}{2} \right)  \qquad \hbox{or} \qquad
\lambda_- = \left(\frac{n}{2},\frac{n}{2} + s \right) \ , 
\ee
where $n=0,1,\ldots$.
It is clear that these representations contain the spin $s$ representation in their diagonal. 
The 
eigenvalue of the tensor harmonics only depends on $\lambda$ (or equivalently $n$), and 
for $\lambda$ of the form (\ref{reps}) is given by \cite{Camporesi:1994ga}
\be{Lapei}
-E^{(s)}_n = 2\left[C_2\left({n\over 2}+s\right)+C_2\left({n\over 2}\right)\right]-C_2(s)
= (s+n) (s+n+2) - s \ ,
\ee
where $C_2(j)=j(j+1)$ is the usual second order Casimir for the $SU(2)$ representation 
labelled by $j$. 

For each such $\lambda$ (or $n$), the label $I$ takes $(n+2s+1) \cdot (n+1)$
different values; for $s>0$ there are then $2 \cdot (n+2s+1) \cdot (n+1)$ different
transverse and traceless rank $s$ tensor harmonics with the same eigenvalue $E_n^{(s)}$,
whereas for $s=0$ (scalar harmonics), the two choices $\lambda_\pm$ coincide,
and the degeneracy is  $(n+1)^2$, as is familiar from the description of 
the hydrogen atom. In the following we shall only be considering the 
transverse and traceless tensor harmonics corresponding to the representations (\ref{reps}). 

To write out (\ref{genhar}) more explicitly, we specify a section as 
\be{gensect}
\sigma(x)=(g_L(x), g_R(x)) \ , \qquad \hbox{where} \quad
g_L(x) \cdot g_R^{-1}(x)=x \ .
\ee
The tensor harmonics for $\gl=\gl_+=(\frac{n}{2}+s,\frac{n}{2})$ 
are then explicitly 
\be{psidef}
\Psi^{(s)(n +; m_1,m_2)}_a(x) = \sum_{k_1,k_2} 
\langle s,a| \frac{n}{2}+s,k_1;\frac{n}{2},k_2\rangle \, 
D^{(\frac{n}{2}+s)}_{k_1,m_1}(g_L^{-1}(x)) \, D^{(\frac{n}{2})}_{k_2,m_2}(g_R^{-1}(x)) \ ,
\ee
while for $\gl=\gl_-=(\frac{n}{2},\frac{n}{2}+s)$ we have instead
\be{psimdef}
\Psi^{(s)(n -; m_1,m_2)}_a(x) = \sum_{k_1,k_2} 
\langle s,a| \frac{n}{2},k_1;\frac{n}{2}+s,k_2\rangle \, 
D^{(\frac{n}{2})}_{k_1,m_1}(g_L^{-1}(x)) \, D^{(\frac{n}{2}+s)}_{k_2,m_2}(g_R^{-1}(x)) \ .
\ee
In either case $I=(m_1,m_2)$ labels the different states in $\lambda$ and thus denotes
different tensor harmonics. Concentrating for definiteness on $\lambda=\lambda_+$, 
$\langle s,a| \frac{n}{2}+s,k_1;\frac{n}{2},k_2 \rangle$ is the Clebsch-Gordon coefficient
describing the decomposition of the tensor product 
$(\frac{n}{2}+s)\otimes (\frac{n}{2})$ into spin $s$, while 
$D^{(j)}_{m,n}(g)$ is  the $(m,n)$-matrix element of the $SU(2)$ rotation $g$ in the 
representation $j$. 
The above wavefunctions are normalized so that 
\be{norma1}
\sum_a \int d\mu(x)\, \Psi_a^{(s)(n +;m_1,m_2)}(x)^\ast \, 
\Psi_a^{(s)(n +;m_1',m_2')}(x) = \frac{2\pi^2 (2s+1)}{(n+2s+1)(n+1)}\, \delta^{m_1,m_1'}\,
\delta^{m_2,m_2'}\ ,
\ee
where $d\mu(x)$ is the Haar measure on $S^3=SU(2)$, normalized so that
the volume of $S^3$ is $2\pi^2$.

\subsection{Choice of Section}

The formula (\ref{psidef}) (or (\ref{psimdef})) 
obviously depends on the choice of a section 
$\sigma(x)$ or, in other words, of $(g_L(x), g_R(x))$. We will now concentrate,
for reasons that will become clearer later, on two out of infinitely many choices of sections. 

The first, which we call the `canonical section' is in some sense the most obvious choice:
\be{can}
\sigma_{can}(x) = (g_L(x),g_R(x))=(e,x^{-1}) \ .
\ee
With respect to the induced basis of tensor functions, the tensor harmonics
labelled by  $(n; m_1,m_2)$ in (\ref{psidef}) are given as 
\begin{eqnarray}
\Psi^{(s)(n +; m_1,m_2)}_{a(can)}(x) & = & \sum_{l_1,l_2} 
\langle s,a| \frac{n}{2}+s,l_1;\frac{n}{2},l_2 \rangle \, 
D^{(\frac{n}{2}+s)}_{l_1,m_1}(e) \, D^{(\frac{n}{2})}_{l_2,m_2}(x)  \nonumber \\
& = & \langle s,a| \frac{n}{2}+s, m_1;\frac{n}{2},a-m_1 \rangle \, 
D^{(\frac{n}{2})}_{a-m_1,m_2}(x) \ . \label{cansec}
\end{eqnarray}
This answer is simple in some respects,
being given purely in terms of single $SU(2)$ rotation matrix elements. 

The second choice of section we will consider is a so-called
`thermal section' because it respects the thermal quotient symmetry. 
As we shall explain in more detail below, the thermal quotient is obtained by
the group action 
\be{groupa}
x \mapsto  A \, x \, B^{-1} \ , 
\ee
where $A$ and $B$ are fixed elements of $SU(2)$. Given any such
group action, there are special sections that respect this symmetry. By this one means 
that the quotient acts on the principal bundle $G=SU(2)\times SU(2)$ in a way which 
commutes with the {\it right} action by $H=SU(2)$. This is achieved by having the
quotient act by a {\it left} action on $G$. Not all sections of the principal bundle will be 
compatible with this left action in the sense of obeying
\be{specsec}
\left(g_L(AxB^{-1}), g_R(AxB^{-1})\right)= \sigma\left(A\, x\, B^{-1} \right) = 
(A,B) \cdot \sigma(x) = (A\cdot g_L(x), B\cdot g_R(x))  \ .
\ee
The thermal section will turn out to obey this relation in the case of thermal quotients. 

In terms of the spherical coordinates of (\ref{sphercoords})
and (\ref{spheref1}), a thermal section is given by
\be{geone1}
g_L(\chi, \theta,\phi) = \left(
\begin{array}{cc}
\cos\frac{\theta}{2} \, e^{i(\phi+\chi)/2}  \quad 
& - \sin\frac{\theta}{2} \, e^{i(\phi-\chi)/2} \cr
\sin\frac{\theta}{2} \, e^{-i(\phi-\chi)/2} \quad 
& \cos\frac{\theta}{2} \, e^{-i(\phi+\chi)/2} 
\end{array} \right) \ , 
\ee
and 
\be{getwo1}
g_R(\chi, \theta,\phi)= \left(
\begin{array}{cc}
\cos\frac{\theta}{2} \, e^{i(\phi-\chi)/2}  \quad
&  - \sin\frac{\theta}{2} \, e^{i(\phi+\chi)/2}  \\
\sin\frac{\theta}{2} \, e^{-i(\phi+\chi)/2} \quad
&  \cos\frac{\theta}{2} \, e^{-i(\phi-\chi)/2} 
\end{array}\right) \ .
\ee
For the following it will be important that these group elements factorize as 
\be{facto}
g_L(x)=U(\hat{n})e^{i{\chi\over 2}\sigma_3} \ , \qquad 
g_R(x)=U(\hat{n})e^{-i{\chi\over 2}\sigma_3} \ ,
\ee
where $\sigma_3$ is the usual Pauli matrix
\be{uhatn}
\sigma_3 = \left(\begin{array}{cc} 1 & 0 \\ 0 & -1 \end{array}\right) \ , \qquad
\hbox{and} \qquad 
U(\hat{n}) = \left(
\begin{array}{cc}
\cos\frac{\theta}{2} e^{i\frac{\phi}{2}} & - \sin\frac{\theta}{2} e^{i\frac{\phi}{2}}  \\
\sin\frac{\theta}{2} e^{-i\frac{\phi}{2}}  & \cos\frac{\theta}{2} e^{-i\frac{\phi}{2}} 
\end{array} \right) \ .
\ee
Note that $U(\hat{n})$ can be viewed as a (local) section for the principal 
$U(1)$ (Hopf) bundle $S^3$ over the base $S^2$. This section is well defined 
except at the poles $\theta=0,\pi$. 

Later on we shall also need the thermal section in the double polar coordinates
 (\ref{torcoords}), for which it takes the form
\begin{equation}\label{gLtrot}
g_L (\gp,\eta ,\varphi)= \left(
\begin{array}{cc}
e^{i(\varphi-\eta)/2} \cos\frac{\gp}{2} \quad & 
ie^{i(\varphi-\eta)/2} \sin\frac{\gp}{2}  \\[8pt]
ie^{-i(\varphi-\eta)/2} \sin\frac{\gp}{2} \quad &
e^{-i(\varphi-\eta)/2} \cos\frac{\gp}{2} 
\end{array} \right) \ ,
\ee
and 
\begin{equation}\label{gRtrot}
g_R(\gp, \eta, \varphi) = \left(
\begin{array}{cc}
 e^{i(\varphi + \eta)/2} \cos\frac{\gp}{2} \quad & 
  -ie^{i(\varphi+\eta)/2} \sin\frac{\gp}{2}  \\[8pt]
  -ie^{-i(\varphi + \eta)/2} \sin\frac{\gp}{2} \quad &
 e^{-i(\varphi+ \eta)/2} \cos\frac{\gp}{2} 
\end{array} \right) \ .
\ee
Note that in these coordinates we can write 
\be{torfi1}
g_L (\gp,\eta ,\varphi) = e^{i\frac{(\varphi-\eta)}{2} \sigma_3} V(\psi) \ , \qquad
g_R (\gp,\eta ,\varphi) = e^{i\frac{(\varphi+\eta)}{2} \sigma_3} V(\psi)^{-1} \ , 
\ee
where
\be{Vdef}
V(\psi) = \left( 
\begin{array}{cc}
\cos\frac{\psi}{2} & i \sin\frac{\psi}{2} \\
 i \sin\frac{\psi}{2} & \cos\frac{\psi}{2}
 \end{array} \right) \ .
 \ee
It is straightforward to check that with both sets of coordinates we have indeed 
$g_L(x)g_R^{-1}(x)=x$, where $x$ is of the form (\ref{spheref1}) and
(\ref{torusf}), respectively. 
The expression for the components of the tensor harmonics are then given by 
(\ref{psidef}) with $g_L(x)$, $g_R(x)$ as above. There is no immediate simplification
(see however section~\ref{otherbasis} below), and the expressions are more
complicated than (\ref{cansec}). 

\subsection{Explicit Formulae}

In order to illustrate the general construction from above 
we shall now exhibit some explicit solutions. This will also
allow us to connect our formulae to existing results in the literature. 
The reader who is not interested in this detailed comparison 
may proceed directly to Sec.~3.

\subsubsection{The Scalar Case}

The scalar case ($s=0$) is the simplest since the answer will be 
independent of the choice of section, as we shall verify momentarily.
In fact, using the general formula (\ref{psidef}) for $s=a=0$ we get (recall
that $\lambda_+=\lambda_-$ in this case)
\be{sc}
\Psi^{(n;m_1,m_2)}(x) = \sum_{m}  \, 
\langle 0,0| \frac{n}{2}, -m ; \frac{n}{2}, m \rangle \, 
D^{(\frac{n}{2})}_{-m,m_1}\bigl(g_L(x)^{-1}\bigr) \, 
D^{(\frac{n}{2})}_{m, m_2}(g_R(x)^{-1}) \ ,
\ee
where $g_L(x)$ and $g_R(x)$ are any section, {\it i.e.}\ satisfy 
$g_L(x) \cdot g_R(x)^{-1} = x$. Using
\be{CG1}
\langle 0,0| \frac{n}{2}, -m;\frac{n}{2},m \rangle = \frac{(-1)^{\frac{n}{2}-m}}{\sqrt{n+1}} \ ,
\ee
as well as the fact that 
$D^{(j)}_{-m,m_1}(g_L^{-1}) =(-1)^{m_1+m}D^{(j)}_{-m_1,m}(g_L)$ we 
can do the sum over $m$ in (\ref{sc}) explicitly, and we obtain
\be{sc1}
\Psi^{(n;m_1,m_2)}(x) = \frac{(-1)^{\frac{n}{2} + m_1} }{\sqrt{n+1}}\, 
D^{(\frac{n}{2})}_{-m_1,m_2}(x)  \ .
\ee
This is evidently independent of the chosen section. All these functions have eigenvalue 
$\gl_n= -n(n+2)$. Since $m_1,m_2$ each range over $(n+1)$ values, we have a total 
degeneracy of $(n+1)^2$. The answer (\ref{sc1}) is also familiar from the 
Peter-Weyl theorem as forming a complete, orthonormal basis for functions on $S^3$.

\subsubsection{Factorization}\label{otherbasis}

In the spherical coordinates of (\ref{sphercoords}) the sphere $S^3$ is parametrized 
in terms of the angles $(\theta,\phi)$ defining an $S^2$, times a radial coordinate 
$\chi$. Typical results for tensor harmonics available in the literature 
({\it e.g.}\ \cite{Camporesi:1994ga, Camporesi:1995fb}) are usually 
given in a factorized form in terms of these coordinates. 
However, our group theoretic basis of eigenfunctions \eq{psidef}, \eq{psimdef} with the 
thermal section \eq{geone1}, \eq{getwo1} does not exhibit such a factorization. 
To compare with the results in the literature we will consider particular 
linear combinations of the 
group theoretic eigenfunctions which exhibit this factorization.

For example, for the scalar harmonics, we define
\be{scalguess}
\Phi_{n\, l\, m}(x) = \frac{n+1}{\sqrt{2\pi^2}}\, 
\sum_{m_1,m_2}\langle \frac{n}{2}, m_1 ; \frac{n}{2}, m_2 | l,m \rangle 
\Psi^{(n; m_1,m_2)}(x) \ ,
\ee
where $l=0, 1\ldots n$, and $m$ runs over the $(2l+1)$ values $m=-l\ldots l$, thus
accounting again for the $(n+1)^2$ fold degeneracy of the scalar harmonics with
eigenvalue 
\begin{equation}
\Delta_{(0)} \Phi_{n\, l\, m} = - n (n+2) \Phi_{n\, l\, m} \ .
\end{equation}
A straightforward computation then exhibits the factorized form
\be{eigfn}
\Phi_{n\, l\, m}(\chi, \theta , \phi) =C_{n \, l} 
\frac{1}{(\sin\chi)^{1/2}} P^{-l -1/2}_{n + 1/2} (\cos\chi)  \, 
Y^{lm} (\theta, \phi) \ ,
\ee
were $C_{n\, l} = \sqrt{(n+1){(n+l+1)!\over (n-l)!}}$ and 
$P^{-l -1/2}_{n+ 1/2}$ is the associated Legendre function of the
first kind, which can be 
expressed either in terms of hypergeometric functions or Jacobi Polynomials 
(see \cite{GR})
\bea{relhyp}
P^{-l -1/2}_{n + 1/2} (\cos\chi) &=& \frac{1}{\Gamma( l + 3/2)}
\left( \frac{\sin \chi/2}{\cos\chi/2} \right)^{l+1/2} 
F( -n -1/2, n + 3/2, l + 3/2; \sin^2 \chi/2) \cr
&=& 2^{-l-\hf}{(n-l)!\over \GG(n+\thf)}\sin^{l+\hf}{\gk}\, P_{n-l}^{(l+\hf,l+\hf)}(\cos{\gk}) \ .
\eea
$Y^{lm}(\theta, \phi)$ are the normalized (scalar) spherical harmonics on 
$S^2$.

\medskip

\noindent For the general case of spin $s$, we define, using the thermal section,
\be{newbasis}
\begin{array}{rcl}
{\displaystyle \Phi^{+ (s)}_{a, n\, l\, m}(\gk,\theta,\phi) }& = &   
{\displaystyle \sum_{m_1,m_2}
\langle \frac{n}{2}+s, m_1 ; \frac{n}{2}, m_2 | l,m\rangle 
\Psi^{(s)(n +; m_1,m_2)}_{a(therm)}(x) } \nonumber \vspace*{0.1cm}\\
{\displaystyle \Phi^{- (s)}_{a, n\, l\, m}(\gk,\theta,\phi)}
 & = &   
{\displaystyle \sum_{m_1,m_2}
\langle \frac{n}{2}, m_1 ; \frac{n}{2}+s, m_2 | l,m\rangle 
\Psi^{(s)(n -; m_1,m_2)}_{a(therm)}(x)  \ , }
\end{array}
\ee
where $l$ runs over the values 
\be{ldef}
l=s,s+1,\ldots,s+n \ ,
\ee
while $m$ takes the $(2l+1)$ values $m=-l,-l+1,\ldots,l-1,l$; altogether we 
thus have again
\be{count}
2 \cdot \sum_{l=s}^{s+n} (2l+1) = 2 (n+1) (2s+n+1)
\ee
different solutions. To see that these solutions are again in factorized form we 
insert the definition of $\Psi^{(s)(n \pm; m_1,m_2)}_{a(therm)}(x)$ from (\ref{psidef})
and (\ref{psimdef}) 
into (\ref{newbasis}), and use (\ref{facto}) as well as (\ref{d1}). A straightforward 
computation then shows that  
\be{fact1}
\Phi^{\pm (s)}_{a, n\, l\, m}(\gk,\theta,\phi)  = Q^{\pm(s)}_{a,n\, l}(\chi) \, D^{(l)}_{a,m}
(U^\dagger(\hat{n})) \ , 
\ee
where
\be{Qdef}
\begin{array}{rcl}
Q^{+(s)}_{a, n\, l}(\gk) & = & 
{\displaystyle \sum_k  \langle s ,a| \frac{n}{2}+s,k;\frac{n}{2},a-k\rangle \, 
e^{-i\chi(2k-a)} 
\langle \frac{n}{2}+s, k ; \frac{n}{2}, a-k |l,a\rangle  }\\
Q^{-(s)}_{a, n\, l}(\gk) & = & 
{\displaystyle \sum_k  \langle s ,a| \frac{n}{2},k;\frac{n}{2}+s,a-k\rangle \, 
e^{-i\chi(2k-a)} \langle \frac{n}{2}, k ; \frac{n}{2}+s, a-k |l,a\rangle \ .}
\end{array}
\ee
Since $U(\hat{n})$ is only a function of $(\theta,\phi)$, (\ref{fact1}) thus
gives a formula for the harmonics in factorized form. In fact, the $D^{(l)}_{a,m}
(U^\dagger(\hat{n}))$ are equivariant functions on $S^2$ under the $U(1)$ action of the principal 
$U(1)$ bundle over $S^2$. Thus they correspond to different tensor harmonics on $S^2$. They are 
the same as the usual spin-weighted spherical harmonics of Newman and Penrose, 
and essentially the same as the familiar monopole harmonics \cite{Dray:1984gy}. 

For the spinor case, 
$s=\half$, we have checked that the resulting harmonics agree precisely
with the explicit formulae given in \cite{Camporesi:1995fb}. Actually, these
functions are also eigenfunctions of the Dirac operator 
$\sldel$ with eigenvalues $\pm i(n+\thf)$, and thus 
the eigenvalue with respect to $\sldel^2$ is $-(n+\thf)^2$. This differs from 
$E_n^{(1/2)}$ in (\ref{Lapei}) by a constant (independent of $n$)
whose origin lies in the non-trivial curvature of $S^3$. 

We have also worked out (\ref{newbasis}) for the vector harmonics
$s=1$, and compared them to the explicit formulae of \cite{Camporesi:1994ga}. 
In identifying these solutions with each other one has to take into account,
as mentioned in section~2.1, that the components of the harmonics in the thermal
section are defined with respect to the standard vielbein on $S^3$, see
eq.\ (\ref{vielc}). On the other hand, the vector harmonics of \cite{Camporesi:1994ga}
are given with respect to a coordinate basis. It follows from (\ref{vielc}) that the
dictionary between the two bases is 
\be{basch}
\Psi_{\pm 1} = {1\over \sqrt{2}\sin{\chi}}
\Bigl[{1\over \sin{\gt} }\Psi_{\phi}\mp i\Psi_{\gt} \Bigr]\ , \qquad \Psi_0=\Psi_{\gk}\ ,
\ee
where we have suppressed the $[\pm, (n,l,m)]$ labels that are common on both
sides. Once this is taken into
account, the above group theory solutions $\Phi^{\pm(1)}_{a,n\, l\, m}$ agree
precisely with (linear combinations) of the harmonics given in \cite{Camporesi:1994ga}.

\section{Heat Kernel on $S^3$}

With this detailed understanding of the spin $s$ harmonics we can
now calculate the spin $s$ heat kernel as per (\ref{heatkdef})
\be{heat1}
K^{(s)}_{ab}(x,y;t) = 
\sum_{(n \pm ;m_1,m_2)} a_n^{(s)}\, 
\Psi^{(s)(n \pm ;m_1,m_2)}_a (x) \, 
\Bigl( \Psi^{(s)(n \pm ;m_1,m_2)}_b (y)\Bigr)^\ast  \, e^{E_n^{(s)} t} \ ,
\ee
where $x$ and $y$ are two points of $S^3$, and the sum runs over all
spin $s$ harmonics labelled by $(n\pm ;m_1,m_2)$ as above. Furthermore, 
$E_n^{(s)}$ is defined in (\ref{Lapei}), while 
the normalisation constant $a_n^{(s)}$ equals
\be{ans}
a_n^{(s)} = \frac{1}{2\pi^2}\, \frac{(n+2s+1) \, (n+1)}{(2s+1)} \ .
\ee
This normalizes the heat kernel so that, using (\ref{norma1}), we get 
\be{inthe}
\sum_a \int d\mu(x)\, K_{aa}^{(s)} (x,x;t) = \sum_{n=0}^{\infty} d_n^{(s)} \, e^{E_n^{(s)} t} \ ,
\ee
where 
\be{ddef}
d_n^{(s)} = (2 - \delta_{s,0}) \, (n+1) \, (n+2s+1) 
\ee
is the total multiplicity of transverse spinor harmonics of eigenvalue
$E_n^{(s)}$. (The prefactor $(2-\delta_{s,0})$ takes into account that 
for $s>0$ there are two sets of harmonics for each $n$, while for $s=0$ 
there is only one.) Note that (\ref{inthe}) is the `trace' over the heat kernel
that is important for the calculation of the one-loop determinant.
\smallskip

Inserting our general formula for the harmonics, see eq.~(\ref{psidef}), the heat kernel
becomes
\bea{heat2}
K^{(s)}_{ab}(x,y;t) & =  &
\sum_{l_1,l_2;m_1,m_2} \, \sum_{p_1,p_2;q_1,q_2} a_n^{(s)}\, 
\langle s,a| l_1, p_1;l_2, p_2 \rangle \, \langle l_1, q_1; l_2 , q_2 | s,b\rangle \, 
 e^{E_n^{(s)} t}  \nonumber \\
& & \quad \qquad\qquad\times 
D^{(l_1)}_{p_1,m_1}(g_L(x)^{-1}) \,
\Bigl( D^{(l_1)}_{q_1,m_1}(g_L(y)^{-1}) \Bigr)^\ast \nonumber \\
& & \quad \qquad \qquad \times 
D^{(l_2)}_{p_2,m_2}(g_R(x)^{-1}) \,
\Bigl( D^{(l_2)}_{q_2,m_2}(g_R(y)^{-1}) \Bigr)^\ast \ , 
\eea
where $(l_1,l_2)$ runs over all pairs of representations of the form 
$(\frac{n}{2}+s,\frac{n}{2})$ or $(\frac{n}{2},\frac{n}{2}+s)$, and $E_n^{(s)}$,
expressed in terms of $(l_1,l_2)$, equals
\be{End}
E_n^{(s)} = -(s+n) (s+n+2) + s = - 2 \Bigl[l_1(l_1+1) + l_2 (l_2+1) \Bigr] + s(s+1)   \ .
\ee
Since the representations are unitary we have
\begin{equation}
\Bigl(D^{(l_1)}_{q_1,m_1}(g_L(y)^{-1}) \Bigr)^\ast = D^{(l_1)}_{m_1,q_1}(g_L(y)) \ , \qquad
\Bigl(D^{(l_2)}_{q_2,m_2}(g_R(y)^{-1}) \Bigr)^\ast = 
D^{(l_2)}_{m_2,q_2}(g_R(y)) \  .
\end{equation}
Thus we can perform the sum over $m_1$ and $m_2$ and obtain
\bea{heat21}
K^{(s)}_{ab}(x,y;t) & =  &
\sum_{l_1,l_2} \, \sum_{p_1,p_2;q_1,q_2} a_n^{(s)}\, 
\langle s,a| l_1, p_1;l_2, p_2 \rangle \, \langle l_1, q_1; l_2 , q_2 | s,b\rangle \,   
e^{E_n^{(s)} t}  \nonumber \\
& & \quad \times \,
D^{(l_1)}_{p_1,q_1} \Bigl(g_L(x)^{-1} g_L(y)\Bigr) \, 
D^{(l_2)}_{p_2, q_2} \Bigl(g_R(x)^{-1} g_R(y)\Bigr) \ .
\eea
Written in terms of the more abstract description of the tensor harmonics, 
eq.\ (\ref{genhar1}), this formula takes the form
\be{heat211}
K^{(s)}_{ab}(x,y;t) = 
\sum_{\lambda} a_n^{(s)} \, U^{\lambda}(\sigma(x)^{-1} \sigma(y))_{ab} \, 
e^{E_n^{(s)} t} \ ,
\ee
where $\lambda$ runs over all the representations of the form (\ref{reps}), and
$a_n^{(s)}$ and $E_n^{(s)}$ are as defined in (\ref{ans}) and (\ref{Lapei}), respectively. 
Furthermore, the matrix elements are taken in the spin $s$ subrepresentation with
respect to the diagonal $SU(2)$. 
Finally, we can also use (\ref{d1}) to rewrite (\ref{heat21}) as 
\bea{heat3}
K^{(s)}_{ab}(x,y;t) & =  &
\sum_{a',b'} \,  D^{(s)}_{aa'}(g_L(x)^{-1}) \,  D^{(s)}_{b'b}(g_L(y)) \\
& &  \hspace{-0.2cm} \times
\sum_{l_1,l_2}\, \sum_{p_1,p_2;q_2} a_n^{(s)}
\langle s,a'| l_1, p_1;l_2, p_2 \rangle \, \langle l_1, p_1; l_2 , q_2 | s,b' \rangle \,  
  e^{E_n^{(s)} t}  
D^{(l_2)}_{p_2,q_2} (x\, y^{-1}) \ , \nonumber 
\eea
where we have used that $g_L(y) g_R(y)^{-1} = y$ and similarly for $x$.

The un-integrated heat kernel (\ref{heat21}) and (\ref{heat3}) obviously depends in general 
on the choice of section, as is clear, for instance, from the first line of (\ref{heat3}). 
Indeed this dependence just reflects the way the components of the harmonics themselves 
depend on  the choice of section, see (\ref{A2}). For the case of the scalar, this ambiguity is 
not present and one can write the final answer explicitly, which we do in the next subsection. 
For higher spin, the expression cannot be simplified further unless one makes a specific choice 
of section (as also coordinates). We exhibit the answer for the thermal section in Sec.~3.2.

\subsection{The Scalar Case}

In the scalar case, $s=0$, the representation labels $a$ and $b$ are trivial, 
and so is the first line of (\ref{heat3}). The scalar heat kernel is then of the form 
\bea{heats}
K^{(0)}(x,y;t) & =  & \frac{1}{2\pi^2}  \, 
\sum_{n=0}^{\infty}\, \sum_{m} (n+1)^2 
|\langle \frac{n}{2}, m; \frac{n}{2} ,-m  | 0,0\rangle |^2 \, 
e^{- n (n+2) t}   \, D^{(\frac{n}{2})}_{m,m} (y\, x^{-1})  \nonumber  \\
& =  & \frac{1}{2\pi^2} \,
\sum_{n=0}^{\infty}\,  (n+1) e^{- n (n+2) t}\, \Tr_{\frac{n}{2}}(y\, x^{-1}) \ ,
\eea
where we have used (\ref{CG1}).  Since
\be{tra}
\Tr_{\frac{n}{2}}(y\, x^{-1}) = {\sin(n+1)\rho \over \sin{\rho}} \ ,
\ee
where $\rho$ is the geodesic distance between $x$ and $y$, we can rewrite
the scalar heat kernel as 
\be{scalk}
K^{(0)}(\rho;t)={1\over 2\pi^2}\sum_{n=0}^{\infty}
 (n+1) {\sin(n+1)\rho\over \sin{\rho}}e^{-n(n+2)t} \ .
\ee
This reproduces the answer given, for example, in 
\cite{Camporesi:1990wm}.

\subsection{Higher Spin}

As mentioned above, for larger $s$, (\ref{heat3}) does not simplify further, unless 
we make some specific choices. In the following we shall use the 
spherical coordinates (\ref{sphercoords}), and consider the 
thermal section (\ref{geone1}) and (\ref{getwo1}). 

Since $S^3$ is a homogeneous space, we may, without loss of generality, 
assume the point $y$ to be at the `origin', {\it i.e.}\ to be  represented by the identity 
matrix 
\begin{equation}
g_L(y)=g_R(y)=e \ .
\end{equation}
The thermal section for the other point $x$ is then described by (\ref{facto}).
Then we can write (\ref{heat21}) as 
\bea{heatherm}
K^{(s)}_{ab}(x,e;t) & =  &
\sum_{l_1,l_2} \, \sum_{p_1,p_2;q_1,q_2} a_n^{(s)}\, 
\langle s,a| l_1, p_1;l_2, p_2 \rangle \, \langle l_1, q_1; l_2 , q_2 | s,b\rangle \,   e^{E_n^{(s)} t}  \nonumber \\
& & \quad \times \,
D^{(l_1)}_{p_1,q_1} \Bigl( e^{-i{\chi\over 2}\sigma_3} U^{\dagger}(\hat{n}) \Bigr) \, 
D^{(l_2)}_{p_2, q_2} \Bigl(e^{i{\chi\over 2}\sigma_3} U^{\dagger}(\hat{n}) \Bigr) \nonumber \\
& =& \sum_{l_1,l_2} \, \sum_{p_1,p_2;q_1,q_2} a_n^{(s)}\, 
\langle s,a| l_1, p_1;l_2, p_2 \rangle \, \langle l_1, q_1; l_2 , q_2 | s,b\rangle \,   e^{E_n^{(s)} t}  \nonumber \\
& & \quad \times \,
e^{i(p_2-p_1)\chi}D^{(l_1)}_{p_1,q_1}( U^{\dagger}(\hat{n}) ) \, 
D^{(l_2)}_{p_2, q_2} (U^{\dagger}(\hat{n}) ) \nonumber \\
& =& \sum_{b'}D^{(s)}_{b'b} (U^{\dagger}(\hat{n}) ) \sum_{l_1,l_2} \, a_n^{(s)}\, e^{E_n^{(s)} t}  
\nonumber \\
& & \quad \times \,
\sum_{p_1, p_2}  
\langle s,a| l_1, p_1;l_2, p_2 \rangle \, \langle l_1, p_1; l_2 , p_2 | s,b' \rangle \,
e^{i(p_2-p_1)\chi}
\nonumber \\
& \equiv & D^{(s)}_{ab} (U^{\dagger}(\hat{n}) )  \, K^{(s)}_{a}(\chi, 0;t)\ .
\eea
In the penultimate line we have employed the identity (\ref{d1}), and in 
the last line we have used that the Clebsch Gordan coefficents vanish unless
$b'=a$. Finally, we have defined 
\bea{heat22}
K^{(s)}_{a}(\chi,0;t) & =  & 
\sum_{l_1,l_2} \, \sum_{p_1,p_2} a_n^{(s)}\, 
| \langle l_1, p_1; l_2 , p_2 | s,a\rangle|^2 \, 
e^{E_n^{(s)} t}  \, e^{i\chi(p_2-p_1)} \  . 
\eea
We should mention in passing that this form of the heat kernel in spherical coordinates 
can also be deduced from the alternative factorized form of the 
eigenfunctions $\Phi^{\pm(s)}_{a, nlm}$ that we obtained in (\ref{fact1}). 

The radial part of the heat kernel $K^{(s)}_{a}(\chi,0;t)$ can be evaluated 
using the explicit form of the Clebsch-Gordan coefficents appearing in 
(\ref{heat22}); this is carried out in Appendix C. 
The final answer is 
\be{spinsheatkdiagfin1}
K^{(s)}_{a}(\chi,0;t)= \frac{1}{2 \pi^2}\, \frac{1}{(2s+1)}\, 
\sum_{n=0}^{\infty} 
{(n+1)! (2s+1)!\over (n+2s)!}  \, K^{(s)}_{a;n}(\chi)\, e^{E_n^{(s)}t}\ ,
\ee
where $K^{(s)}_{a;n}(\chi)$ is given in terms of Gegenbauer polynomials in 
(\ref{spinssumk3}). It follows from the explicit formula for $K^{(s)}_{a;n}(\chi)$ that 
\begin{equation}
K^{(s)}_{a;n}(\chi=0)= (2-\delta_{s,0})\, {(n+2s+1)!\over n! (2s+1)!}\ ,
\end{equation}
and thus for $\chi=0$ the complete heat kernel simplifies to 
\begin{equation}
K^{(s)}_{ab}((\chi=0,\theta,\phi),e;t) = 
 D^{(s)}_{ab} (U^{\dagger}(\hat{n}) )  \, 
\frac{1}{2 \pi^2}\, \frac{1}{(2s+1)}\, \sum_{n=0}^{\infty} 
d_n^{(s)}\, e^{E_n^{(s)} t} \ , 
\end{equation}
where $U(\hat{n})$ was defined in terms of $(\theta,\phi)$ in (\ref{uhatn}), and the 
mutliplicity $d_n^{(s)}$ was introducted in (\ref{ddef}).

\subsubsection{The Spinor Case}

As a cross check we can compare with some of the existing results in the literature. 
We have already evaluated the scalar case. The next simplest case is then the 
spinor case $(s=\frac{1}{2})$. This has been obtained explicitly in, for instance 
\cite{Camporesi:1995fb}. The only small difference is that they evaluate the heat 
kernel for the operator $\sldel^2$ rather than the spinor Laplacian. The eigenvalues 
of the former are $-(n+\thf)^2$ while that of the latter are $-(n+\thf)^2+\thf$. Taking 
this shift into account, the result given there (see {\it e.g.}\ eq.~(3.4) of  the published version of 
\cite{Camporesi:1995fb} or eq.~(4.12) of the {\tt arXiv} version) is
\be{spink}
K^{(\frac{1}{2})}_{ab}((\chi,0,0),e;t) = 
\delta_{ab} \, \Bigl[ \frac{1}{2 \pi^2} \sum_{n=0}^{\infty} (n+1)(n+2) \phi_n(\chi) \, 
e^{-t(n+\frac{3}{2})^2+\thf t} \Bigr]\ , 
\ee
where $\phi_n(\chi)$ is given in terms of Jacobi polynomials as
\be{phjac}
\phi_n(\chi)  = 
{n!\, \GG(\thf)\over \GG(n+\thf)}\cos{\chi\over 2}\, P^{(\hf ,\thf)}_n(\cos{\chi})\ .
\ee
Using the recursion
\be{jacrec}
P_n^{(\hf, \thf)}(\cos{\chi}) =
P_n^{(\hf,\hf)}(\cos{\chi})
-\sin^2{\chi\over 2}P_{n+}^{(\thf, \thf)}(\cos{\gk})\ ,
\ee
and the relation of the Jacobi polynomials $P_n^{(m,m)}$ to the Gegenbauer 
polynomials we find  
\begin{equation}
\phi_n(\chi)={2\over (n+1)(n+2)}\cos{\chi\over 2}
\Bigl[C_n^2(\cos\chi)-C_{n-1}^2(\cos\chi)\Bigr]\ .
\end{equation}
Putting this back in (\ref{spink}), we find that it agrees precisely with the general expression 
in (\ref{spinsheatkdiagfin1}) for the special case of $(s=\hf)$.

\subsubsection{The Vector Case}

As a last example we write the answer for the vector case ($s=1$) in full detail. 
We again consider the heat kernel for the points between the north pole $e$, and the point
$(\chi, \theta, \phi)$ on $S^3$. 
The heat kernel is obtained from 
(\ref{heatherm}) and (\ref{spinsheatkdiagfin1}) 
\begin{equation}
\label{vec-ker1}
K^{(1)}_{ab}((\chi,\theta,\phi), e ; t) = 
D_{ab}^{(1)}(U^\dagger(\hat{n}))\,  {1\over \pi^2}\sum_{n=0}^{\infty}{1\over (n+2)} K^{(1)}_{a;n}(\chi) \, 
e^{-t ((n+1)(n+3)-1)} \ ,
\end{equation}
and (\ref{spinssumk3}) implies that the explicit expressions for $K^{(1)}_{a;n}(\chi)$ are 
\be{Kerv1}
K^{(1)}_{1;n}(\chi) = K^{(1)}_{-1;n}(\chi) = 2 \, \left[
\cos\chi \, C_n^3(\cos\chi) -2\, C_{n-1}(\cos\chi) +\cos\chi \, C_{n-2}^3(\cos\chi) \right] \ ,
\ee
and
\be{Kerv0}
K^{(1)}_{0;n}(\chi)=2\, C_n^2(\cos\chi) \ .
\ee
It is also useful to rewrite this expressions in terms of trignometric functions. Using 
(\ref{Geg}) and the recursion relations satisfied by the Gegenbauer polynomials 
we find that
\begin{eqnarray}
\label{vec-trig}
K^{(1)}_{0;n}(\chi) &=& 
 \frac{1}{2\sin^3\chi} \Bigl( ( n+3) \sin( n+1)\chi - (n+1) \sin( n+3)\chi \Bigr)\ , \\
\nonumber
K^{(1)}_{1;n}(\chi) &=& K^{(1)}_{-1;n}(\chi) 
 =- \frac{1}{8 \sin^3\chi} \Bigl[ ( 2+n)(3+n) \sin n\chi 
 - 2 ( 2+ 4n + n^2) \sin( n+2) \chi  \\ \nonumber
& & \qquad\qquad \quad\; \; \;
+ (n+2) (n+1) \sin( n+4)\chi \Bigr] \ .
\end{eqnarray}
The above form of the radial heat kernel is suitable for analytical continuation 
to AdS$_3$ (see section \ref{h3-vector}).

\section{Heat Kernel on Thermal $S^3$}

In perparation for the calculation on thermal $H_3^+$ we now 
want to study the heat kernel on the thermal quotient of $S^3$, {\it i.e.}\
on the manifold $S^3/\Gamma$, where $\Gamma$ describes a specific group 
of identifications. These identifications are most easily described in the 
double polar coordinates (\ref{torcoords}), where the action of the generator 
$\gamma$ of  $\Gamma$, is given by 
\be{therm1}
\gamma: \quad \eta \mapsto \eta + \beta \ , \qquad
\varphi \mapsto \varphi + \vartheta \ . 
\ee
In order for this group action to be globally well-defined, we should take 
$\Gamma$ to be of finite order, $\Gamma\cong {\mathbb Z}_N$, {\it i.e.}\ 
$\gamma^N=1$. This corresponds to a Lens space quotient of $S^3$. 
The generator $\gamma$ acts on the group element $g$ in (\ref{torusf}) as 
\be{gact}
g \mapsto \tilde{g} = 
\left( \begin{array}{cc}
e^{i\frac{\tau}{2}} & 0 \cr 
0 & e^{-i\frac{\tau}{2}}  
\end{array} \right) \, g \, 
\left( \begin{array}{cc}
e^{-i\frac{\bar\tau}{2}} & 0 \cr 
0 & e^{i\frac{\bar\tau}{2}}  
\end{array} \right)  = A \, g \, \bar{A}^{-1}\ , 
\ee
where 
\be{taudef}
\tau\equiv \tau_1-\tau_2 =\vartheta-\beta ; \, \quad \bar\tau \equiv \tau_1+\tau_2 = \vartheta + \beta
\ee
and 
\begin{equation}
A = \left( \begin{array}{cc}
e^{i\frac{\tau}{2}} & 0 \cr 
0 & e^{-i\frac{\tau}{2}}  
\end{array} \right) \ , \qquad
\bar{A} = \left( \begin{array}{cc}
e^{i\frac{\bar\tau}{2}} & 0 \cr 
0 & e^{-i\frac{\bar\tau}{2}}  
\end{array} \right) \ .
\end{equation}
The section that is compatible with this group action must satisfy 
(compare (\ref{specsec}))
\be{specsec1}
\sigma(\gamma(x)) = (A,\bar{A}) \cdot \sigma(x) \ .
\ee
As explained above (\ref{specsec}), such a choice of section is necessary for the 
compatibility of the thermal quotient with the coset space identification on the 
principal bundle $G$. 
Another way to understand this requirement is as follows. The group action
(\ref{gact}) induces a natural map (via push forward) relating the tangent basis
at $g$ to that at $\tilde{g}$. On the other hand, the choice of 
section specifies a vielbein (see (\ref{sigmaten})) for all $g\in G$. The
condition (\ref{specsec}) implies that the vielbein at $\tilde{g}$ agrees precisely
with the push-forward via (\ref{gact}) of the vielbein at $g$.

Obviously, (\ref{specsec1}) is not satisfied by every section; in particular, it
is not true for the `canonical' section (\ref{can}). On the other hand, one easily checks 
that it is satisfied by the thermal section (\ref{gLtrot}) and (\ref{gRtrot}). 

\subsection{Method of Images}

The heat kernel on the quotient space can be calculated from that on $S^3$ by the method 
of images.  We can fix one of the points (say $x$) and sum over the images of the second 
one ($y$). This is to say, we have
\be{thermalheat}
\sum_{m\in {\mathbb Z}_N}\,  \, 
 K^{(s)}_{ab}
(x,\gamma^m(y);t) \ ,
\ee
where $N$ is the order of $\gamma$.
We will be interested in obtaining the determinant of $\Delta_{(s)}$ on $S^3/ \Gamma$, which
means that we need to find the integrated traced heat kernel for coincident points on the 
orbifolded space, {\it i.e.}
\be{orbheat}
\sum_{m\in {\mathbb Z}_N}\,  \sum_a \int_{S^3/\Gamma} d\mu(x)\, 
 K^{(s)}_{aa}
(x,\gamma^m(x);t) \ .
\ee
Here we have traced over the group theory indices $a,b$ with a simple Kronecker delta 
since we are working in a tangent space basis (such as the usual vielbein basis for 
the $s=1$ case). If we were working in a coordinate basis, then the expression would be 
more complicated, involving a Jacobian factor such as 
${\partial \gamma(x)^{\mu} \over \partial x^{\nu}}$  \cite{Giombi:2008vd}. 

Since we are considering the identification \eq{therm1}, we need to understand the heat kernel evaluated
at two points $x$ and $y=\gamma^m(x)$ that have the same value
for the $\psi$-component (and only differ in their $\eta$- and 
$\varphi$-component). In this case it follows from (\ref{torfi1}) that 
\be{si1}
g_L(x)^{-1} g_L(y) = V(\psi)^{-1} \, U_1  \, V(\psi) \ , \qquad
g_R(x)^{-1} g_R(y) = V(\psi) \, U_2 \, V(\psi)^{-1} \ ,
\ee
where $V(\psi)$ is defined in (\ref{Vdef}) with $\psi=\psi(x)=\psi(y)$, 
and $U_1$ and $U_2$ are of the form
\be{si2}
U_1 = \exp\Bigl(  i\frac{ (\Delta\varphi - \Delta\eta)}{2} \sigma_3\Bigr) = e^{ i m \frac{\tau}{2} \sigma_3} \ , \quad
U_2 = \exp\Bigl(  i \frac{(\Delta\varphi + \Delta\eta)}{2} \sigma_3 \Bigr) 
= e^{i m\frac{\bar\tau}{2} \sigma_3}
\ee
with $\Delta\varphi =\varphi(y)-\varphi(x)=m\vartheta$ and 
$\Delta\eta =\eta(y)-\eta(x)=m\beta$, and additionally using the definition \eq{taudef}. 
With these conventions (\ref{heat21}) for the particular case of $y=\gamma^m(x)$ becomes
\bea{heat4}
K^{(s)}_{ab}(x,y=\gamma^m(x) ;t) & =  &
\sum_{l_1,l_2} \, \sum_{p_1,p_2;q_1,q_2} a_n^{(s)}\, 
\langle s,a| l_1, p_1;l_2, p_2 \rangle \, 
\langle l_1, q_1; l_2 , q_2 | s,b\rangle \, 
e^{E_n^{(s)} t}  \nonumber \\
& & \quad \times \,
D^{(l_1)}_{p_1,q_1} \Bigl(V(\psi)^{-1} U_1 V(\psi))\Bigr) \, 
D^{(l_2)}_{p_2,q_2} \Bigl(V(\psi) \, U_2 V(\psi)^{-1} \Bigr) \ . \nonumber
\eea
We can write the trace over $a=b$ more abstractly as 
\begin{eqnarray}
\label{heat51}
& & \sum_a K^{(s)}_{aa}(x,\gamma^m(x);t)  \nonumber \\
& & \quad =   
\sum_{l_1,l_2} a_n^{(s)}\, e^{E_n^{(s)} t} \, 
\Tr_{s} \Bigl[ \Bigl(V(\psi)^{-1} U_1 V(\psi) \Bigr)^{(l_1)} \otimes 
\Bigl(V(\psi) \, U_2 V(\psi)^{-1} \Bigr)^{(l_2)} \Bigr] \ ,
\end{eqnarray}
where the trace is only taken over the spin $s$ subrepresentation in the tensor
product $(l_1\otimes l_2)$. Conjugation with the operator 
$V(\psi) \otimes V(\psi)$ does not modify the trace (since the subpresentation
$s$ is invariant under the action of $g\otimes g$), and thus (\ref{heat51}) can be 
rewritten as 
\be{heat61}
\sum_a K^{(s)}_{aa}(x,y;t)  =   
\sum_{l_1,l_2} a_n^{(s)}\, e^{E_n^{(s)} t} \, 
\Tr_{s} \Bigl[  U_1^{(l_1)} \otimes 
\left(V(\psi)^2 U_2 V(\psi)^{-2} \right)^{(l_2)} \Bigr] \ .
\ee
Let us denote a general diagonal group element by
\be{Dalp}
D(\alpha) = \left( \begin{array}{cc}
e^{i\alpha} & 0 \\ 0 & e^{-i\alpha} 
\end{array} \right) \ .
\ee
Since both $U_1$ and $U_2$ are diagonal, it follows that 
\be{si}
D(\alpha) U_1 D(\alpha)^{-1} = U_1 \ , \qquad
D(\beta) U_2 D(\beta)^{-1} = U_2 \ .
\ee
Taking $\alpha = -(\varphi-\eta)/2$ and $\beta = - (\varphi+\eta)/2$, 
and using the same argument as in going to (\ref{heat61}), we then obtain
\be{heat71}
\sum_a K^{(s)}_{aa}(x,y;t)  =   
\sum_{l_1,l_2} a_n^{(s)}\, e^{E_n^{(s)} t} \, 
\Tr_{s} \Bigl[  U_1^{(l_1)} \otimes 
\left( g\, U_2\, g^{-1} \right)^{(l_2)} \Bigr] \ ,
\ee
where 
\begin{equation}
g = D\Bigl((\varphi-\eta)/2\Bigr) V(\psi)^2 D\Bigl(-(\varphi+\eta)/2\Bigr)  
= \left(
\begin{array}{cc}
e^{-i\eta} \cos\psi \quad & i e^{i\varphi} \sin\psi \\
i e^{-i\varphi} \sin\psi  \quad & e^{i\eta} \cos\psi
\end{array}
\right) = g(\psi, \eta, \varphi)  \ ,
\end{equation}
and $g(\psi, \eta, \varphi)$ is defined in (\ref{torusf}). 
Next we perform the integral over $S^3/\Gamma$ in (\ref{orbheat}). 
This amounts to integrating (\ref{heat71}) over $\psi$ in the fundamental
domain of $S^3/\Gamma$. Equivalently, we may integrate $\psi$ over
the full range $\psi\in [0,\frac{\pi}{2}]$, and divide by the appropriate volume factor.
In addition, since (\ref{heat71}) is
actually independent of $\eta$ and $\varphi$ --- this is obvious from
(\ref{heat61}) --- we may also integrate $\eta,\varphi\in[0,2\pi]$. But then
the second group element in (\ref{heat71}) equals
\begin{equation}
\int_{S^3} dg \left( g\, U_2 \, g^{-1}\right)^{(l_2)}  = 
\frac{2\pi^2}{\dim(l_2)} \, 
\Tr_{(l_2)}  (U_2) \, {\bf 1}_{l_2} \ ,
\end{equation}
where we have used Schur's lemma, observing that the operator on the left hand side
commutes with all group elements. Thus the integrated heat kernel becomes
\begin{eqnarray}
& & \int_{S^3/\Gamma} d\mu(x) \sum_a K^{(s)}_{aa}(x,\gamma^m(x);t)   \nonumber \\
& & \qquad \qquad = 
\pi \tau_2\, 
\sum_{l_1,l_2} \frac{a_n^{(s)}}{\dim(l_2)} \, \Tr_{(l_2)}(U_2) 
e^{E_n^{(s)} t} \, 
\Tr_{s} \Bigl[  U_1^{(l_1)} \otimes  {\bf 1}^{(l_2)} \Bigr] \ , \label{heat8}
\end{eqnarray}
where the prefactor $\pi \tau_2 = 2 \pi^2 \frac{\tau_2}{2\pi}$ comes from the relative volume
of $S^3/\Gamma$ to $S^3$. The final trace can now be easily done
(for example using similar arguments as above), and
it equals
\be{tr1}
\Tr_{s} \Bigl[  U_1^{(l_1)} \otimes  {\bf 1}^{(l_2)} \Bigr]  = \Tr_{(l_1)}(U_1) \,
\frac{2s+1}{\dim(l_1)} \ .
\ee
 Plugging this back into (\ref{heat8}) we therefore obtain
 \begin{equation}
 \int_{S^3/\Gamma}d\mu(x)
  \sum_a K^{(s)}_{aa}(x,\gamma^m(x);t) =  \frac{\pi \tau_2}{2\pi^2}\,
 \sum_{l_1,l_2}  \Tr_{(l_1)}(U_1) \,  \Tr_{(l_2)} (U_2)  \, e^{E_n^{(s)} t} \ ,
 \end{equation}
 where we have used the formula for $a_n^{(s)}$ from (\ref{ans}). 
 Finally, doing the sum over $m$ leads to 
\begin{eqnarray}
& & \sum_{m\in {\mathbb Z}_N} \sum_a \int_{S^3/\Gamma}d\mu(x) 
K^{(s)}_{aa}(x,\gamma^m(x);t)  \nonumber \\
& & \qquad \qquad = {\tau_2\over 2\pi}\, 
\sum_{m\in {\mathbb Z}_N} \sum_{n=0}^{\infty}\Bigl[\chi_{({n\over 2})}(m\tau)
\chi_{({n\over 2}+s)}(m\bar{\tau})+\chi_{({n\over 2}+s)}(m\tau)
\chi_{({n\over 2})}(m\bar{\tau})
 \Bigr]\, 
e^{E_n^{(s)} t}
\nonumber \\
& & \qquad \qquad \equiv  K^{(s)}(\tau,\bar{\tau}, t) \ , \label{25}
\end{eqnarray}
where we have assumed that $s>0$; otherwise the second term in the middle line of 
(\ref{25}) is absent. We have also used the notation
\begin{equation}
\chi_{(l)}(\tau)= 
{\rm Tr}_{(l)}(e^{i \frac{\tau}{2} \sigma_3}) = 
{\sin{(2l+1)\tau\over 2}\over \sin{\tau\over 2}}
\end{equation}
for the $SU(2)$ character in the representation $l$.

\section{Heat Kernel on AdS$_3$}

Having derived the heat kernel for an arbitrary tensor Laplacian on $S^3$ as well as on 
its `thermal' quotient, we will now extend the analysis to the case of $H_3^+$; the thermal
quotient of $H_3^+$ will be discussed in the next section. As mentioned in the 
introduction, this is simplest done by performing a suitable analytic continuation to 
 $H_3^+$  (and its thermal quotient). Since this is, in general, a tricky procedure 
we will motivate and describe in some detail how it is to be carried out. As will become clear, 
for the particular case of $H_3^+$, the central ingredients in our calculation 
(such as the eigenfunctions, eigenvalues and their measure) have been
independently computed and checked to obey the analytic 
continuation from their $S^3$ counterparts, see in particular the series
of papers by Camporesi and Higuchi 
\cite{Camporesi:JGP,Camporesi:1994ga,Camporesi:1995fb}. These 
explicit results can  be taken as the ultimate justification for our use of the analytic continuation 
procedure.

\subsection{Preliminaries}

Euclidean AdS$_3$ is the hyperbolic space $H_3^+$ which can be thought of as the 
homogeneous space
\be{H3}
H_3^+ \cong SL(2,{\mathbb C})/SU(2) \ ,
\ee
where the quotienting is done by the usual right action. We can view  $SL(2,{\mathbb C})$ 
as an  analytic continuation of $SU(2)\times SU(2)$ in a way which will be made explicit 
below.

As in the case of $S^3$ we will need to choose coordinates for explicit expressions. 
Corresponding to the spherical coordinates on $S^3$  \eq{sphercoords} we have now
\be{metrich3}
ds^2 = dy^2 + \sinh y^2 \, ( d\theta^2 + \sin^2 \theta \, d\phi^2) \ ,
\ee
which is obtained by the continuation
$\chi \rightarrow - i y$  and $ ds^2\rightarrow - ds^2$, of \eq{sphercoords}. 

The coset space representative of $SL(2,{\mathbb C})/SU(2)$ (for a given $(y,\theta,\phi)$) 
can be taken to be the continuation of \eq{spheref1} 
\be{h3coset}
\tg(y,\theta,\phi) = \left(
\begin{array}{cc}
\cosh y + \sinh y \, \cos \theta & 
\sinh y \, \sin \theta \, e^{i\phi} \cr
 \sinh y \, \sin \theta \, e^{-i\phi} & \cosh y -  \sinh y\, \cos \theta 
\end{array}
\right) \ .
\ee

For the thermal quotient it will be convenient to work in the double polar coordinate 
analogue of \eq{torcoords}, {\it i.e.}\ to use the metric
\be{torus}
ds^2 = d\rho^2 + \cosh^2 \rho\,  (dt)^2 + \sinh^2\rho\,  (d\varphi)^2 \ .
\ee
This is related to \eq{torcoords} by the continuation 
$\psi \rightarrow - i\rho$, $\eta \rightarrow  i t$ 
and $ ds^2\rightarrow - ds^2$. Therefore 
corresponding to \eq{torusf} we now have the coset space element 
\be{h3torusgroup}
\tg(\rho,t,\varphi) = \left( 
\begin{array}{cc}
e^{t} \cosh\rho \quad & 
e^{i\varphi}\, \sinh\rho \cr
e^{-i\varphi}\, \sinh\rho \quad 
& e^{-t} \cosh\rho
 \end{array}
 \right) \ .
\ee

To carry through the construction of eigenfunctions as described in Sec.~2, 
we will first need an appropriate  choice of section. As is familiar from the analysis 
of the Lorentz group in four dimensions, the
representations of $SL(2,{\mathbb C})$ are most easily described in terms
of  $SU(2)\times SU(2)$. The Lie algebra of the former is a complexified 
version of the latter. More precisely, if we write the Lie algebra of $SO(4)$ as 
$so(4) \simeq su(2)\oplus su(2)$ with generators $a^{(1)}$ and $a^{(2)}$, respectively, then 
the diagonal $SU(2)$ by which we quotient $SO(4)$ to obtain $S^3$ is generated by
$h=a^{(1)}+a^{(2)}$. Defining $k=a^{(1)}-a^{(2)}$, the complexification 
$k \to -ik$ describes then the continuation from $S^3$ to $H_3^+$. 
This is equivalent to the continuation $\chi \to -iy$ described above. 
  
Thus it will still be useful
to describe the coset representative of $SL(2,{\mathbb C})/SU(2)$ in terms of 
pairs of group elements $(\tg_L,\tg_R)$ that live in the appropriately complexified
version of $SU(2)\times SU(2)$.
The relevant expressions for the complexification are 
obtained from those on $S^3$ precisely by the analytic continuation of the coordinates 
described above. In particular,
the analogue of the thermal section is now described by 
$(\tg_L(x), \tg_R(x))$, where in spherical coordinates we have 
(compare with \eq{geone1} and \eq{getwo1})
\be{glads}
\tg_L(y , \theta,\phi) = \left(
\begin{array}{cc}
\cos\frac{\theta}{2} \, e^{i(\phi-iy)/2}  \quad 
& - \sin\frac{\theta}{2} \, e^{i(\phi+iy)/2} \cr
\sin\frac{\theta}{2} \, e^{-i(\phi+iy)/2} \quad 
& \cos\frac{\theta}{2} \, e^{-i(\phi -iy)/2} 
\end{array} \right) \, = \, U(\hat{n})e^{{y\over 2}\sigma_3}
\ee
and 
\be{grads}
\tg_R(y, \theta,\phi)= \left(
\begin{array}{cc}
\cos\frac{\theta}{2} \, e^{i(\phi+iy)/2}  \quad
&  - \sin\frac{\theta}{2} \, e^{i(\phi -iy)/2}  \\
\sin\frac{\theta}{2} \, e^{-i(\phi-iy)/2} \quad
&  \cos\frac{\theta}{2} \, e^{-i(\phi+iy)/2} 
\end{array}\right) \,  = \, U(\hat{n})e^{-{y\over 2}\sigma_3}.
\ee
In the double polar coordinates which we use for the quotienting, we have similarly 
(compare with \eq{gLtrot} and \eq{gRtrot})
\be{gLt}
\tg_L(\rho, t, \varphi) = \left(
\begin{array}{cc}
{\displaystyle e^{t/2} e^{i\varphi/2} \cosh\frac{\rho}{2}} \quad & 
{\displaystyle  e^{t/2} e^{i\varphi/2} \sinh\frac{\rho}{2} } \\[8pt]
{\displaystyle  e^{-t/2} e^{-i\varphi/2} \sinh\frac{\rho}{2}} \quad &
{\displaystyle e^{-t/2} e^{-i\varphi/2} \cosh\frac{\rho}{2} }
\end{array} \right)
\ee
and 
\be{gRt}
\tg_R(\rho,t, \varphi) = \left(
\begin{array}{cc}
{\displaystyle e^{-t/2} e^{i\varphi/2} \cosh\frac{\rho}{2}} \quad & 
{\displaystyle  - e^{-t/2} e^{i\varphi/2} \sinh\frac{\rho}{2} } \\[8pt]
{\displaystyle  - e^{t/2} e^{-i\varphi/2} \sinh\frac{\rho}{2}} \quad &
{\displaystyle e^{t/2} e^{-i\varphi/2} \cosh\frac{\rho}{2} }
\end{array} \right) \ .
\ee
One can check that with both sets of coordinates we have indeed 
$\tg_L(x)\cdot \tg_R^{-1}(x)=\tg(x)$, where $\tg(x)$ is given in \eq{h3coset} and
\eq{h3torusgroup}, respectively.

\subsection{Harmonic Analysis on $H_3^+$}

As was described in Sec.2.1, to obtain the eigenfunctions of the Laplacian $\Delta_{(s)}$ on $G/H$, 
we need facts from the harmonic analysis on $G$. For a general noncompact semi-simple $G$
 this is an intricate subject (see {\it e.g.}\ \cite{Knapp}). 
However, the results for $G=SL(2,{\mathbb C})$ are relatively well known to 
physicists since $SL(2,{\mathbb C})$ is the Lorentz group in four dimensions. 
Some useful general references on the subject, particularly for the infinite 
dimensional representations which we will need below, are 
\cite{Ruhl,Carmeli}. 
 
The component eigenfunctions of the tensor harmonics are given in terms of matrix 
elements of appropriate unitary representations of  $SL(2,{\mathbb C})$. 
One of the major differences between the compact and the noncompact cases is that 
the (nontrivial) unitary representations of the latter are necessarily infinite dimensional. 
Recall that the usual finite dimensional (and hence non-unitary) 
representations of $SL(2,{\mathbb C})$ are labelled by $(j_1,j_2)$, where
$j_1$ and $j_2$ are the half-integer spin representations of the two $SU(2)$s.
In fact, the most general representation (or the `complete series') of $SL(2,{\mathbb C})$, 
including the unitary representations, can also be labelled by $(j_1,j_2)$, where $j_1,j_2$ 
are now complex but subject to some constraints such as $(j_i-j_2)$ being a half integer. 

The unitary representations come in two series: the so-called `principal series' and the 
`complementary series'. However, only the principal series will play a role in what follows. 
This is because they are the only representations that arise in the decomposition of 
functions on $SL(2,{\mathbb C})$ and therefore (see the discussion around \eq{equivariant})
for sections of bundles on $SL(2,{\mathbb C})/SU(2)$.\footnote{In 
general, additional (normalizable) representations - the `discrete series' - could also 
appear when considering even dimensional hyperbolic spaces.}
These correspond to $j_1$ and $j_2$ taking the values
\be{principal}
2j_1 = s -1+ i \lambda  \ , \qquad
2j_2 = -s - 1+ i \lambda \ ,
\ee
where $\lambda\in{\mathbb R^+}$ and $s$ is half-integer, see
for example \cite[section II.4]{Knapp}. 
When restricted to the diagonal $SU(2)$ subgroup, these representations decompose 
into an infinite number of $SU(2)$ representations of spin $s, s+1, s+2, \ldots$ 
\cite{Ruhl, Carmeli}. Thus these representations play the role of the representations 
$({n\over 2}+s, {n\over 2})$ in the $S^3$ case and will describe the transverse, 
traceless spin $s$ tensors on $H_3^+$. Comparison to \eq{principal} suggests
that the appropriate analytic continuation for $n$ is \cite{Camporesi:1994ga}
\be{analytic}
n\mapsto -s-1+i\lambda \ .
\ee
Thus eigenfunctions of $\Delta_{(s)}$ are given (in the thermal section) by the matrix elements of the
$SL(2,{\mathbb C})$ element
$(\tg_L(x), \tg_R(x))$ in these representations 
labelled by a continuous parameter $\lambda\in{\mathbb R^+}$ (for fixed $s$). 
Their eigenvalues are, up to a sign, given by the same analytic continuation (\ref{analytic})
applied to \eq{Lapei},
\be{hlapei}
E_{\lambda}^{(s)}= - (\lambda^2+s+1) \ .
\ee
The sign is a consequence of the fact that the metric has  changed sign
under the analytic continuation, $ds^2 \rightarrow - ds^2$. Thus the analytic continuation 
of   \eq{Lapei} gives minus the eigenvalue of the Laplacian on $H_3^+$. 

\subsection{The Heat Kernel on $H_3^+$}

In computing the heat kernel the sum over $n$ in (\ref{heatkdef}) is now to be
replaced by an integral over $\lambda$. The measure for the integration 
is determined from the so-called Plancherel measure 
which describes the decomposition of the space of functions on $G$ into its irreducible representations. We will continue to refer to the measure thus obtained for the decomposition of the sections on $G/H$
with spin $s$ (in the case of $G=SL(2,{\mathbb C})$ and $H=SU(2)$) as the Plancherel measure 
and denote it by $d\mu^{(s)}(\lambda)$. 

This Plancherel measure for $H_3^+$ 
(or more generally, the hyperbolic spaces $H_N$) has been 
computed by Camporesi and Higuchi (see for example \cite{Camporesi:JGP, Camporesi:1994ga}). 
The explicit expression is given by 
\be{measure} 
d\mu^{(s)}(\lambda) =  \frac{1}{2\pi^2}\, (2-\delta_{s,0})\, 
\frac{(\lambda^2 + s^2)}{(2s+1)} \, d\lambda \ ,
\ee
which is, up to a sign and the prefactor $(2-\delta_{s,0})$, 
precisely the analytic continuation of the $S^3$ normalisation 
constant $a_n^{(s)} = \frac{1}{2\pi^2} \frac{(n+2s+1)  (n+1)}{(2s+1)}$  (see \eq{ans})
by our analytic continuation (\ref{analytic}). (The origin of this sign is again
the change of sign in the analytic continuation of the metric $ds^2 \rightarrow - ds^2$.
The origin of the prefactor $(2-\delta_{s,0})$ is also the same as before, namely 
that there are two choices $\lambda_\pm$ for $s>0$ (see \eq{reps}), which fall together
for $s=0$.)

The $H_3^+$ heat kernel for spin $s$ fields then
takes the form 
\be{heatkgen}
K^{(s)}_{ab}(x,y;t)= \int_0^{\infty}d\mu^{(s)}(\lambda)\, 
\phi^{(s)}_{\lambda, ab}(x,y)\, e^{-t(\lambda^2+s+1)}\ ,
\ee
where $\phi^{(s)}_{\lambda, ab}(x,y)=U^{\lambda,s}_{ab}(\sigma(x)^{-1}\sigma(y))$ are the 
matrix elements of the representation $(\lambda, s)$ projected onto the spin $s$ representation of 
the diagonal $SU(2)$ ({\it cf.}\ \eq{heat211}). In particular, the index $a$ still labels the components 
of the spin $s$ field and takes values from $-s$ to $s$. The functions $U^{\lambda,s}_{ab}(g)$
are sometimes known as generalised spherical functions (for spin $s$) and have 
many important properties. For example, they are determined completely by knowing 
the values on a maximal torus\footnote{We can decompose a general 
$SL(2,{\mathbb C})$ element $g$ as $g=h_1t\, h_2$, where $h_1,h_2\in SU(2)$ and 
$t$ lies in the maximal torus.}. In spherical polar coordinates this is the statement 
that we know the complete answer to the heat kernel 
once we know the value for one of the points at the 
origin and the other at some $(\chi,0,0)$ for $S^3$ ({\it cf.}\ \eq{heatherm}) and $(y,0,0)$ for $H_3^+$. 
The spherical functions also satisfy simple radial Laplacian equations, which ensures 
that we can also have a simple analytic continuation for them. We refer the reader to 
Sec.~5.3 of \cite{Camporesi:1995fb} for more properties of these spherical functions.

For our purposes it is sufficient to make the following observations.
In the thermal section, using the spherical coordinates \eq{metrich3}, we can 
use a similar reasoning as in  Sec.~3.2. We can choose 
one point to be at the origin and factor out 
the $S^2$ angular dependence as in \eq{heatherm}. Then the other point can be 
taken to be $(y,0,0)$ and we obtain 
\be{heatkgen1}
K^{(s)}_a(y,0;t)= \int_0^{\infty}d\mu^{(s)}(\lambda)\, 
\phi^{(s)}_{\lambda, a}(y)\, e^{-t(\lambda^2+s+1)}\ ,
\ee
where $\phi^{(s)}_{\lambda, a}(y)$ is the analytic continuation of $K^{(s)}_{a;n}(\chi)$ in 
(\ref{spinsheatkdiagfin1}) under $\chi \to -iy$. These functions are expressed in terms of 
Gegenbauer  polynomials in \eq{spinssumk3}.  In order to perform the analytic 
continuation  explicitly,  we can use the definition of the Gegenbauer polynomials 
in terms of hypergeometric functions
\be{geghgeom}
C_n^{\alpha}(\cos\chi)={\Gamma(2\alpha+n)\over \Gamma(n+1)
\Gamma(2\alpha)}\, F\left(2\alpha+n, -n, \alpha+\half; \sin^2{\chi\over 2}\right) \ .
\ee
The right hand side can be defined for complex values of the arguments and in particular 
under the continuation $n\rightarrow -s-1+i\lambda$. 
Note that the index $\alpha$ takes the 
values $s+a+1$ in \eq{spinssumk3} and therefore continues to be an integer. Also the sum 
there continues to be a finite one with an upper limit given by $2a$. 
It is not easy to perform the integral over $\lambda$ for general spin and give an explicit 
form of the heat kernel on AdS$_3$. However, we can do this integral for a few simple
cases and check that the above prescription gives the correct result.

\subsubsection{The Scalar Case}

The heat kernel for the case $s=0$ can be easily evaluated. In this case, we can in 
fact write the answer slightly more generally, namely directly in terms of the geodesic 
separation $r$ between the two points. Instead of \eq{heatherm} we can start with the 
expression \eq{scalk}. Since the metric $ds^2\rightarrow -ds^2$ 
in the analytic continuation we continue $\rho \rightarrow -ir$. Together with 
the continuation $n \rightarrow -1+i\lambda$, we find that \eq{scalk} becomes
\be{kerh3}
K^{(0)}(r; t)
= \frac{1}{2\pi^2} \int_0^\infty d\lambda \, \lambda \, e^{- t( \lambda^2 +1) } \, 
\frac{\sin\lambda r }{ \sinh r} \ ,
\ee
where we have absorbed a sign into the $\lambda$ measure, see
(\ref{measure}). After integrating over $\lambda$ we obtain
\be{fkerh3}
K^{(0)}(r ; t) =\frac{e^{-t}}{( 4 \pi t )^{3/2} } 
\frac{ r \, e^{-\frac{r^2}{4t}} } {\sinh r}  \ .
\ee
The explicit form of the geodesic distance on $H_3^+$ between the points $(y,\theta,\phi)$ and 
$(y',\theta ',\phi ')$ is given by 
\be{defr}
\cosh r = \cosh y'\cosh y -\sinh y'\sinh y\cos\theta'\cos\theta
-\sinh y' \sinh y \sin\theta\sin\theta'\cos(\phi'-\phi) \ .
\ee
The expression \eq{fkerh3} agrees with the heat kernel 
determined in 
\cite{Giombi:2008vd} for the case $m^2=0$ --- the general case is easily 
obtained from this since the mass only contributes an additive term 
to the exponent in \eq{fkerh3}. 

\subsubsection{The Spinor Case}

For $(s= \half )$ we can again take the answer for the sphere, in this case worked out in 
\eq{spink}, and perform the above analytic continuation. 
Instead of writing it in terms of Gegenbauer polynomials we can directly use, for the analytic 
continuation, the 
hypergeometric form of the Jacobi polynomial appearing in \eq{phjac} 
\be{jachgeom}
P_n^{(\alpha, \beta)}(\cos\chi)={\Gamma(n+\alpha+1)\over \Gamma(n+1)\Gamma(\alpha+1)}\, 
F\left(n+\alpha+\beta+1, -n, \alpha+1; \sin^2{\chi\over 2}\right)\ .
\ee
After the continuation $n\rightarrow -\thf +i\lambda$, \eq{spink} then becomes
\be{halfheatkh3}
K^{(\half)}_{ab}=  \delta_{ab} \, \Bigl[ \frac{1}{2\pi^2} \int_{0}^{\infty}
d\lambda\, (\lambda^2+{1\over 4}) \,\phi_{\lambda}(y) \, 
e^{-t(\lambda^2+\frac{3}{2}) } \Bigr]\ , 
\ee
with $\phi_{\lambda}(y)=\cosh{y\over 2} F(\thf+i\lambda, \thf-i\lambda , \thf , -\sinh^2{y\over2})$. 
Here we have again absorbed an overall minus sign into the measure, see
(\ref{measure}). 
This agrees with eq.~(5.14) of \cite{Camporesi:1992tm} (apart from the same shift in 
the exponent, see the discussion before \eq{spink}).

\subsubsection{The Vector Case}
\label{h3-vector}

For $s=1$ we can analytically continue the 
answer for the 3-sphere given in \eq{vec-ker1} and \eq{vec-trig} using 
\begin{equation}
 n\rightarrow -2 + i\lambda\ , \qquad \chi\rightarrow -iy\ .
\end{equation}
For the case where we evaluate the heat kernel between the north pole and the point 
$(y, 0,0 ) = (\chi, 0, 0)$, the geodesic distance $r$ agrees with $y$. Using the above
prescription we then obtain after some straightforward manipulations
\begin{eqnarray}
\label{gt-kernel}
 K_{00}^{(1)}(r, 0;t)  &=& - \sqrt{\frac{\pi}{t}}  \frac{ e^{-2t}}{2\pi^2} 
\left( \frac{1}{\sinh^2 r} e^{-\frac{r^2}{4t}} 
 - \frac{\cosh r }{  \sinh^3 r  }  \int_{0}^r dx e^{ - \frac{x^2}{4t}} \right),  \\ \nonumber 
K_{11}^{(1)}( r, 0;t) &=& K_{-1-1}^{(1)}( r, 0;t), \\ \nonumber 
&=& \frac{e^{-2t}}{4\pi^2 \sin^3 r} \sqrt{\frac{\pi}{t}}
\left( \frac{r}{2t} e^{ - \frac{ r^2}{4t}} \sinh^2 r +  e^{- \frac{r^2}{4t}} \sinh r \cosh r - \int_{0}^r 
dx e^{-\frac{ x^2}{4t}} \right)  \ .
\end{eqnarray}

To check that this result satisfies the heat equation for vectors we recall that the 
heat equation for a $U(1)$ gauge field is given by (see for instance 
\cite{Giombi:2008vd} which we follow by also adding the constant two to the Laplacian)
\begin{equation}
\label{heateqvec}
 -(\Delta_{(1)} +2) K_{\mu\nu'} (x, x'; t) = -\frac{\partial}{\partial t} K_{\mu\nu'}( x, x'; t)\ ,
\end{equation}
where $x = (y, \theta, \phi)$ and $x' = (y', \theta',\phi')$ are two points on $H_3^+$. 
We are interested in the heat kernel satisfying the Lorentz-gauge condition
\begin{equation}
\label{lorentz-g}
 \nabla^\mu K_{\mu\nu'}( x, x'; t) =0\ , \qquad \nabla^{\nu'} K_{\mu\nu'}( x, x'; t) =0\ .
\end{equation}
Thus the initial condition at $t=0$ is 
\begin{equation}
\label{initial}
 K_{\mu\nu'}( x, x'; 0) = g_{\mu\nu'}(x) \, \delta^3(x, x') + 
\nabla_{\mu}\nabla_{\nu'} \frac{1}{\Delta_{(0)}} \delta^3(x, x') \ .
\end{equation}
Since $H_3^+$ is  a maximally symmetric space, we can write the heat kernel, 
following  \cite{Giombi:2008vd}, as 
\begin{equation}\label{coba}
 K_{\mu\nu'}(x, x'; t) = F(t, u) \partial_\mu \partial_{\nu'} u + \partial_\mu \partial_{\nu'} S(t, u) \ ,
\end{equation}
where $1+ u = \cosh r$, and $r$ is the geodesic distance between the points $x$ and $x'$ 
given by (\ref{defr}).  The heat equation (\ref{heateqvec}) then reduces to 
\begin{eqnarray}
 ( \Delta_{(1)} + 1) F(t, u) = \partial_t F(t, u)\ , \\ \nonumber
\Delta_{(1)} S(t, u) - 2 \int_{u}^\infty F(t, v) dv = \partial_t S(t, u) \ ,
\end{eqnarray}
while the Lorentz gauge condition (\ref{lorentz-g}) becomes
\begin{equation}
 \frac{\partial F}{\partial u} (1 +u)  + F  + \partial_t\partial_u S  =0 \ ,
\end{equation}
and the initial conditions on $F$ and $S$ are 
\begin{eqnarray}
 F(0, u) = - \delta^3(x, x') \qquad 
 S( 0, u) = \frac{1}{\Delta_{(0)}}\delta^3( x, x') = - \frac{1}{4\pi} \coth r \ .
\end{eqnarray}
The correct solution is then 
\begin{eqnarray}
\label{soln-lg}
F(r, t)  &=& - \frac{e^{- \frac{r^2}{4t} } }{ (4\pi t)^{3/2} } \frac{r}{\sinh r}\ , \\ \nonumber
S(r, t) & =&  - \frac{2}{( 4\pi)^{3/2} \sqrt{t}} \frac{\cosh r}{\sinh r} \int_{0}^r 
 e^{- \frac{x^2}{4t}}  \ .
\end{eqnarray}
Note that this solution differs form that found in \cite{Giombi:2008vd}, for which
the Lorentz gauge condition was not implemented and which therefore satisfied 
the boundary condtion $K_{\mu\nu'}( x, x';0) = g_{\mu\nu'}(x) \delta^3(x, x')$, which is different 
from (\ref{initial}). In fact, \cite{Giombi:2008vd} had to subtract out a scalar degree of 
freedom from the trace of their heat kernel to obtain the physical one loop determinant
for vectors. This is unnecessary for the solution given in (\ref{soln-lg}) since the Lorentz 
gauge condition guarantees that only the physical degrees of freedom contribute.

In order to compare (\ref{coba}) to (\ref{gt-kernel}) we need to convert the coordinate
basis implicit in  (\ref{coba}) to the tangent space indices of (\ref{gt-kernel}).
For the case where $x$ is the north pole and $x' =( r, 0, 0)$ the relations turn out to be
\begin{eqnarray}
\label{matching}
 K_{00}^{(1)}(r, 0;t) = - F(r, t) \cosh r - \frac{\partial^2  }{\partial r^2} S(r ,t)\ ,  \\ \nonumber
K_{11}^{(1)}(r, 0;t) =K_{-1-1}^{(1)}(r, 0;t)= - F(r, t) - \frac{1}{\sinh r} \frac{ \partial}{ \partial r} S(r, t) \ ,
\end{eqnarray}
where we have used (\ref{basch}). Substituting (\ref{soln-lg}) we then reproduce indeed 
(\ref{gt-kernel}) up to an overall factor of $e^{-2t}$. The origin of this factor is 
that in (\ref{heateqvec}),
following \cite{Giombi:2008vd}, 
we have analyzed the heat equation for $(\Delta_{(1)}+2)$, rather than for the Laplacian
$\Delta_{(1)}$ itself.

\subsection{The Coincident Heat Kernel}

It is difficult to do the integrals over $\lambda$ for the heat kernel in general. However
 it is easy to obtain the expression for the coincident heat kernel for arbitrary spin $s$. 
One need only consider the integrand of \eq{heatkgen} to notice that the coincident traced 
heat kernel $K_{aa}^{(s)}(x,x;t)$ is given by
\begin{eqnarray}
K_{aa}^{(s)}(x,x;t)& = & (2s+1)\, \int_0^{\infty}d\mu^{(s)}(\lambda)\, 
e^{E_{\lambda}^{(s)}t}  \nonumber \\
& = & (2-\delta_{s,0})\,  {1\over 2\pi^2}\int_0^{\infty}d\lambda \, (\lambda^2+s^2)\, 
e^{-t(\lambda^2+s+1)} 
\label{h3coinheat}  \nonumber \\
& = & {1\over (4\pi t)^{\thf}} \, 
(2-\delta_{s,0})\, (1+2s^2 t)\, e^{-t(s+1)} \ . \label{h3coinheatans}
\end{eqnarray}
For $s=1,2$ this agrees precisely with the answers of 
Giombi {\it et.al.}\ \cite{Giombi:2008vd} (up to shifts in the exponent
which come from mass terms), as well as with 
the general expression for the zeta function in 
\cite{Camporesi:1995fb}.

\section{Heat Kernel on Thermal $H_3^+$}

\subsection{The Thermal Identification}

We are actually interested in determining the heat kernel for thermal AdS$_3$. 
Thermal AdS$_3$ is obtained from Euclidean AdS$_3$ ({\it i.e.}\ $H_3^+$) 
described above by
identifying points under a ${\mathbb Z}$ action. To identify the relevant
${\mathbb Z}$ action it is useful to write $H_3^+$ in double polar coordinates \eq{torus},
which were obtained from the corresponding coordinates on $S^3$ by the continuation
\begin{equation}
i\psi =  \rho \ , \qquad i \eta =  - t \ .
\end{equation}
Translating the thermal identifications \eq{therm1} of $S^3$ into the analytically
continued variables then corresponds to 
\begin{equation}
t \sim t - i \beta \ , \qquad \phi \sim \phi + \vartheta \ .
\end{equation}
Thus $\beta$ has the interpretation of the inverse temperature. In addition, 
the analytically continued variables, $\tau$ and $\bar\tau$ of \eq{taudef}
are now
\begin{equation}
\tau = \vartheta + i \beta \ , \qquad \bar\tau = \vartheta - i \beta \ ,
\end{equation}
and are  indeed complex conjugates of one another.

\subsection{The Heat Kernel}

As discussed in Sec.~5, we could analytically continue the harmonic analysis on $S^3$ 
to that on $H_3^+$. We are now considering quotients of these two spaces. 
The identifications being made in the quotienting are also analytic continuations of each other,
as seen in the previous subsection. 
We therefore expect that the expressions for the heat kernel on the thermal quotient of $S^3$ 
described in Sec.~4 should be analytically continued as well. 
However, it should be pointed out that the group $\Gamma\cong {\mathbb Z}_N$ 
generated by $\gamma$ in the $S^3$ case is finite in order for the identifications to make 
global sense. 
There is no such constraint in the case of the identifications on $H_3^+$, and therefore
the group is just ${\mathbb Z}$. This difference however only plays a role when taking into 
account the sum over the images to obtain the full heat kernel: in the thermal $S^3$ case
\eq{thermalheat} is a finite sum, while the corresponding sum for $H_3^+$ (see below) will
involve an infinite sum over $m$.

However, this is a global aspect of the quotienting which we expect to be irrelevant to the 
analytic continuation of a particular image point to the heat kernel. Indeed,
the analysis of section~4.1 was essentially algebraic, and thus can be equally applied for
the case of $H_3^+$. There we had written the expressions in terms of group integrals and 
as traces over the appropriate $SU(2)$ representations.  These group theoretic operations 
carry over into the noncompact case though care should be taken in the group integrals 
and definitions of the trace.
This is normally accomplished through the various ingredients of the harmonic analysis 
on the noncompact groups that we have mentioned so far. The additional feature we 
need to use in our analytic continuation of the results of Sec.~4.1 is the trace. 
For a noncompact group one can define what is called the the 
Harish-Chandra (or global) character which is defined as a distributional analogue 
of the usual trace. In the case of $SL(2, {\mathbb C})$ this has been worked out 
and will be explained more explicitly below.

Using these ingredients we will assume the analysis of Sec.~4.1 can be carried through in 
an identical fashion for $SL(2, {\mathbb C})$; in the following we shall consider, for ease of
notation, the case $s>0$ --- the calculation for $s=0$ is almost identical. 
Instead of the $SU(2)\times SU(2)$ character 
given  in \eq{25} we now end up with a character of the $SL(2, {\mathbb C})$ 
element $M={\rm diag}(e^{i\tau\over 2}, e^{-i\tau\over 2})$. 
The $SL(2, {\mathbb C})$ character for an element with diagonal entries 
$(\alpha,\alpha^{-1})$ is given by 
(see {\it e.g.}\   \cite[p.~100]{Ruhl}  or \cite[p.~117]{Carmeli} --- note that there is a typo
in \cite{Carmeli})
\be{sl2char}
\chi_{(j_1,j_2)}(\alpha)={\alpha^{2j_1+1}\bar \alpha^{2j_2+1}
+\alpha^{-2j_1-1}\bar \alpha^{-2j_2-1}
\over | \alpha-\alpha^{-1}|^2}\ .
\ee
Thus the final answer for the integrated heat kernel for the case of 
thermal $AdS_3$ takes the form ({\it cf.}\ (\ref{25}))
\be{h3orbheat}
K^{(s)}(\tau,\bar{\tau}; t)=2\cdot {\tau_2\over 2 \pi}\, 
\sum_{m\in {\mathbb Z}} \int _{0}^{\infty}d\lambda\,
 \chi_{\lambda,s}(e^{im\tau\over 2})\, e^{-t(\lambda^2+s+1)} 
\ee
with
\be{chisl2}
\chi_{\lambda,s}(e^{im\tau\over 2})= \half {\cos(m s\tau_1- m \lambda \tau_2)\over 
|\sin \frac{m\tau}{2}|^2} \ , 
\ee
which is just the character of $M$ evaluated for $j_1=\half(s-1+i\lambda)$ and
$j_2= \half(-s-1+i\lambda)$. Since $s>0$ we also have to consider the contribution where
the roles of $j_1$ and $j_2$ are interchanged, and this is responsible 
for the overall factor of $2$ in (\ref{h3orbheat}).
For fixed $m$ the integral over 
$\lambda$ of
\begin{equation}
\frac{\tau_2}{2\pi\, |\sin \frac{m\tau}{2}|^2} \int_0^{\infty} 
d\lambda \, \cos(m s\tau_1- m \lambda \tau_2) e^{-t(\lambda^2+s+1)} 
\end{equation}
can be peformed by Gaussian integration, and we obtain 
\begin{equation}
 {\tau_2\over 4\sqrt{\pi t} 
|\sin{m\tau\over 2}|^2}\cos(m s \tau_1)e^{-{m^2 \tau_2^2\over 4 t}}e^{-(s+1)t} \ .
\end{equation}
The term with $m=0$ diverges; it describes the integrated heat kernel 
on $H_3^+$ since for $m=0$ the two points $y=\gamma^m(x) = x$ and
$x$ coincide. The divergence is then simply a consequence of the infinite 
volume of $H_3^+$. In any case, the contribution with $m=0$ is 
independent of $\tau$, and therefore not of primary interest to us. 
Subtracting it out, the final result is then 
\begin{equation}
\label{thermalh3-ker}
K^{(s)}(\tau,\bar{\tau}; t)=\sum_{m=1}^{\infty}{\tau_2\over \sqrt{4\pi t} 
|\sin{m\tau\over 2}|^2}\cos(s m\tau_1)e^{-{m^2\tau_2^2\over 4 t}}e^{-(s+1)t} \ .
\end{equation}
This is the central result of the paper which we shall use extensively below. 
\smallskip

For the case $s=1$, (\ref{thermalh3-ker}) gives exactly the answer of \cite{Giombi:2008vd}
for the transverse components as given in their eqs.~(4.16) and (4.17). 
(Note that their $2\pi\tau$ is our $\tau$; furthermore the relative factor $e^{2t}$ comes
from the curvature contribution in their eq.~(2.15).) For the case of $s=2$, while the 
contribution from the transverse components is not separately considered in  
\cite{Giombi:2008vd}, it can be inferred from their result eq.~(4.25) (together with 
eq.~(4.22)). In fact the first term in their eq.~(4.25) is exactly equal to (\ref{thermalh3-ker}) with
$s=2$ (again up to a relative factor of $e^{2t}$ coming from the curvature contribution).
In the next section we also check that the correct one loop graviton determinant is reproduced 
by this result.

The expression \eq{thermalh3-ker} for the case of $s=0$ and $s=1$ is of the form given by the 
Selberg trace formula for scalars and transverse vectors. In fact, the heat kernel for these cases 
were written down in \cite{Bytsenko:1997sr} using the Selberg trace formula --- see their eqs.~(B.1)
and (B.2). (A general reference for the trace formula in this context is \cite{Bytsenko:1994bc}, 
Sec.~3.4, see also \cite{Bytsenko:2008ex, Bytsenko:2008wj}). 
The trace formula essentially gives a path integral like interpretation to the heat kernel answer. 
To summarize the salient points we note that the 
sum over $m$ is a sum over closed paths of 
non-zero winding number $m$ and of length $m\tau_2$ weighted with a classical action 
${m^2\tau_2^2\over 4t}$. The denominator in \eq{thermalh3-ker} is proportional to
$|1-q^m|^2$ (with $q=e^{i\tau}$). This is the semiclassical (or van-Vleck) determinant. 
Finally, from the explicit form of the $s=1$ case quoted in eq.~(B.1) of \cite{Bytsenko:1997sr}, 
one interprets the $\cos{m\tau_1}$ piece of \eq{thermalh3-ker} as a monodromy term. This suggests 
that the  general spin $s$ answer given by us here can be understood in terms of a general 
Selberg trace formula for symmetric traceless tensors of rank $s$. We should like to mention 
though that the Selberg trace formula is generally applied to quotients of $H_3^+$ of finite volume. 
In such cases there is an additional finite piece coming from the $m=0$ (or `direct') term. 
As mentioned earlier, for the  thermal quotient this is a trivial ($q$ independent) volume divergence.

\section{Partition Function of ${\cal N}=1$ Supergravity}

As an interesting application of the formalism we have developed in the previous
sections we can now evaluate the one loop partition function of 
${\cal N}=1$ supergravity in thermal $H_3^+$ and explicitly check the argument of 
Maloney and Witten \cite{Maloney:2007ud}.  We will, in the process, also derive 
the expressions for the one loop determinant in the bosonic (pure gravity) sector 
reproducing the results of  the check of \cite{Giombi:2008vd}. 

The field content of ${\cal N}=1$ supergravity consists of the graviton of spin $s=2$, 
and the Majorana gravitino of spin $s=3/2$. The complete one loop partition function of ${\cal N}=1$ 
supergravity is therefore the product of the graviton and gravitino contribution
\be{N=1loop}
Z_{1-{\rm loop} } = Z_{1-{\rm loop} }^{\rm graviton} \cdot Z_{1-{\rm loop} }^{\rm gravitino} \ .
\ee
The calculation of the two contributions will be described in detail below, first
for the graviton (Sec.~7.1), and then for the gravitino (Sec.~7.2). In each case
we can reduce the calculation of the one loop 
partition function to determinants of the form $\det ( - \Delta_{(s)} + m_s^2)$, where
$\Delta_{(s)}$ denotes an appropriate spin $s$ Laplacian, while $m_s$ is 
a mass shift. In turn these determinants can be easily deduced from the
heat kernel since we have 
\be{7.1}
- \log {\det ( - \Delta_{(s)} + m_s^2) } = \int_0^\infty \frac{dt}{t}  \, K^{(s)}(\tau,\bar{\tau}; t)  \, 
e^{-m_s^2 t}\ , 
\ee
where $K^{(s)}$ is the spin $s$ heat kernel that was determined above
(\ref{thermalh3-ker}). Thus the knowledge of the heat kernel allows us to calculate
the one loop partition functions fairly directly.

\subsection{ The One Loop Determinant for the Graviton}

The one loop contribution  of the graviton to the effective action has been 
evaluated by several authors 
\cite{Gibbons:1978ji,Christensen:1979iy,Yasuda:1983hk}.  Including the 
gauge fixing terms and the ghosts, the one loop partition function 
for the graviton in $D$ spacetime dimensions is given by \cite{Yasuda:1983hk}
\begin{equation}
 Z_{{\rm 1-loop}}^{{\rm graviton}} = 
{\rm{\det}}^{-1/2} ( \Delta_{(2)}^{\rm LL} - 2 R/D ) \cdot  
{\rm{det}}^{1/2} (\Delta_{(1)}^{\rm LL} - 2 R/D) \ ,
\end{equation}
where $\Delta_{(2)}^{\rm LL}$ and $\Delta_{(1)}^{\rm LL}$ denote the  Lichnerowicz Laplacians on 
rank $2$ symmetric traceless and vectors, respectively, while $R$ is 
the scalar curvature. For $H_3^+$ the curvature tensors, in units of the radius of 
AdS$_3$, are 
\begin{equation}
\label{curv}
R_{\mu\rho\nu\sigma} = \frac{R}{6}  \,  
( g_{\mu\nu}g_{\rho\sigma} - g_{\mu\sigma} g_{\nu\rho} ) \ , \qquad
R_{\mu\nu} = \frac{R}{3} \, g_{\mu\nu} \ , \qquad R=-6 \ . 
\end{equation}
Note that the convention for the scalar curvature used in 
\cite{Yasuda:1983hk} differs by a sign from the above (conventional) definition.

To convert the Lichnerowicz  Laplacian to the ordinary Laplacian we use the 
relations  \cite{Gibbons:1978ji} 
\begin{eqnarray}
\label{defintions}
 \Delta_{(2)}^{\rm LL} \, T_{\mu\nu} &=& - \Delta_{(2)} T_{\mu\nu} - 2 R_{\mu\rho\nu\sigma} 
 T^{\rho\sigma} + R_{\mu\rho} T^\rho{}_\nu +R_{\nu\rho} T_\mu{}^\rho
\\ \nonumber
\Delta_{(1)}^{\rm LL} \, T_\mu &=& - \Delta_{(1)} T_\mu + R_{\mu\rho} T^\rho \ ,
\end{eqnarray}
where $T_{\mu\nu}$ and $T_\mu$ are arbitrary symmetric traceless tensors and
vectors, respectively. Using (\ref{curv}) we then find 
\begin{eqnarray}
\label{shifts}
( \Delta_{(2)}^{\rm LL}  - 2R/D) \,T_{\mu\nu}  &=&  (-\Delta_{(2)} - 2) \, T_{\mu\nu}
  \\ \nonumber
( \Delta_{(1)}^{\rm LL} -  2R/D)\, T_\mu &=&  (-\Delta_{(1)} + 2) \, T_\mu \ .
\end{eqnarray}
Thus the one loop partition function of the graviton is given by 
\begin{equation}
\label{1-loop-grav}
 Z_{{\rm 1-loop}}^{{\rm graviton}} = 
{\rm{\det}}^{-1/2} (- \Delta_{(2)} - 2  )  \cdot {\rm{det}}^{1/2} (-\Delta_{(1)} +2 ) \ ,
\end{equation}
which can be directly evaluated in terms of the heat kernel. 
In fact, using (\ref{7.1}) we simply have 
\begin{eqnarray}
\log   Z_{{\rm 1-loop}}^{{\rm graviton}}  &=  & 
- \frac{1}{2} \log( {\rm det} ( -\Delta_{(2)}  -  2) )  
+ \frac{1}{2} \log ( {\rm det} ( - \Delta_{(1)}  + 2))  \\ \nonumber
&=&
\frac{1}{2}\,  \int_0^\infty \frac{dt}{t} \left( K^{(2)}(\tau, \bar\tau; t) \, e^{2t}
-  K^{(1)}(\tau, \bar\tau; t)  \, e^{-2t} \right) \ .
\end{eqnarray}
Using the expression (\ref{thermalh3-ker}) for the heat kernel, and performing the 
$t$-integral with the help of 
\begin{equation}
\label{integral}
{1\over 4\pi^{1/2}}\int_0^{\infty}{dt\over t^{3/2}}e^{-{\alpha^2\over 4t}-\beta^2 t}
={1\over 2\alpha}e^{-\alpha\beta} \ , 
\end{equation}
we then obtain 
\begin{eqnarray}
\log Z_{\rm 1-loop}^{{\rm graviton} } &=&  \frac{1}{2} \sum_{m=1}^\infty 
\frac{1}{m|\sin \frac{m\tau}{2} |^2 }
\left( \cos( 2m\tau_1) e^{-m\tau_2} - \cos( m\tau_1) e^{-2m\tau_2}\right)  \\ \nonumber
&=& \sum_{m=1}^\infty \frac{1}{m} \left( \frac{q^{2m}}{1- q^m} + \frac{\bar q^{2m}}{ 1-\bar q^m} 
\right)  
= - \sum_{n=2}^\infty \log |1-q^n|^2 \ ,
\end{eqnarray}
where $q = \exp(i\tau)$, and in the last line we have expanded out the geometric series.
Thus the one loop gravity partition function is given by
\begin{equation}\label{graviton-final}
 Z_{{\rm 1-loop}}^{{\rm graviton}} = \prod_{n=2}^\infty \frac{1}{|1- q^n|^2} \ .
\end{equation}
This was argued to be the result for pure gravity in \cite{Maloney:2007ud} by 
a quantum extension of the argument of Brown and Henneaux
\cite{Brown:1986nw}. It also reproduces precisely the calculation of \cite{Giombi:2008vd}.
Including the tree level contribution $|q|^{-2k}$, the total one loop gravity partition function 
is just the product of a left- and a right-moving Virasoro vacuum representation 
at $c=\bar{c} = 24 k$ \cite{Maloney:2007ud}. Since there are no bulk propagating states in 
$3d$ gravity, the perturbative partition function simply counts the contributions of the so-called 
boundary Brown-Henneaux states which are obtained by acting on the $SL(2, {\mathbb C})$ 
invariant vacuum by the Virasoro generators $L_{-n}$ (with $n\geq 2$).

\subsection{One Loop Determinant for the Gravitino}

The calculation for the one loop gravitino partition function is slightly more complicated. 
The gravitino that is of relevance to us is a Majorana gravitino, but it is actually
easier to study first the case of a Dirac gravitino. Its action is given by \cite{Rarita:1941mf}
\begin{equation}
\label{lag}
 S = - 
\int d^3 z \sqrt{g} \, \bar \psi_\mu ( \Gamma^{\mu\nu\rho} 
D_\nu \psi_\rho + \hat m \, \Gamma^{\mu\nu} ) \psi_\nu \ .
\end{equation}
Here  $\Gamma^\mu$ are  defined as $\Gamma^\mu = \gamma^a e^\mu_a$ with 
$e^\mu_a$ being the vielbeins, and 
\begin{equation}
\label{gamma-m}
 \gamma_0 = 
\left( \begin{array}{cc}
0 & -i \\
i & 0 
\end{array} \right)\ , \quad
\gamma_1 = 
\left( \begin{array}{cc}
0 & 1  \\
1 & 0 
\end{array}
\right)\ , \quad
\gamma_2 = 
\left( \begin{array}{cc}
1 & 0 \\
0 & -1 
\end{array}
\right)\ . \quad
\end{equation}
The $\Gamma$-matrices satisfy the usual Clifford algebra, $\{\Gamma^\mu,\Gamma^\nu\}=2 g^{\mu\nu}$, 
and we define
\begin{eqnarray}
 \Gamma^{\mu\nu} &=& \frac{1}{2} ( \Gamma^\mu\Gamma^\nu - \Gamma^\nu\Gamma^\mu) \label{7.14}
 \\ \nonumber
\Gamma^{\mu\nu\rho} &=& \frac{1}{3!} (
 \Gamma^\mu \Gamma^\nu \Gamma^\rho - \Gamma^\nu\Gamma^\mu \Gamma^\rho 
+ \;\mbox{cyclic} ) \ .
\end{eqnarray}
Furthermore the  covariant derivative is given by 
\begin{equation}
 D_\mu \psi_\nu = \partial_\mu\psi_\nu + \frac{1}{8} \omega_\mu^{ab} [\gamma_a, 
 \gamma_b]\psi_\nu 
- \tilde\Gamma^\rho_{\mu\nu} \psi_\rho \ ,
\end{equation}
where $\tilde\Gamma^\rho_{\mu\nu}$ are the Christoffel symbols, while
$\omega_\mu^{ab}$ refers to the spin connection.  
For a massless gravitino
$\hat m$ is  related to the radius of AdS$_3$  by
\begin{equation}
\label{gravi-mass}
 \hat m^2 = \frac{1}{4} \ .
\end{equation}
The gravitino Lagrangian has the gauge symmetry
\begin{equation}
\label{gauge-inv}
 \delta \psi_\mu = D_\mu \epsilon - \hat m \Gamma_\mu \epsilon \ ,
\end{equation}
and thus we need to worry about isolating the gauge invariant degrees of freedom. 
To do so we shall fix a gauge and use the Fadeev-Popov method, following 
\cite{Fradkin:1983nw}. To start with we remove from $\psi_\mu$ the gauge trivial part
\be{ex1}
 \psi_\mu  = \varphi_\mu +  \frac{\Gamma_\mu }{3} \psi \ , 
 \ee
where $\Gamma^\mu \varphi_\mu=0$ and $\psi = \Gamma^\mu \psi_\mu$. The remaining field 
$\varphi_\mu$ we then further decompose as
\begin{equation}
\label{expansion}
\varphi_\mu = \varphi_\mu^\bot + \left( D_\mu - \frac{1}{3}\Gamma_\mu \hat D \right) \xi\ , \qquad
\hbox{where} \qquad 
D^\mu \varphi^\bot_\mu = \Gamma^\mu \varphi^\bot_\mu=0\ .
\end{equation}
Here $\hat D =  \Gamma^\mu D_\mu$, and $D_\mu \psi$ is defined by 
\begin{equation}
 D_\mu \psi = \partial_\mu \psi + \frac{1}{8} \omega_\mu^{ab}[\gamma_a, \gamma_b] \psi \ .
\end{equation}
With respect to this decomposition the gravitino Lagrangian (\ref{lag}) then becomes
(the details are described in appendix~D)
\begin{eqnarray}
\label{action}
 S &=& - \int d^3z  \sqrt{g} \left( 
\bar\varphi^{\bot \mu}( \hat D - \hat{m} ) \varphi_\mu^\bot 
- \frac{2}{9} \bar\xi( \hat D - 3 \hat{m})
[ \Delta_{(1/2)}  - 3/4] \xi
\right.  \\ \nonumber
& & \qquad\qquad\qquad  \left. + \frac{2}{9}  \bar \xi \, 
[\Delta_{(1/2)} -3/4] \, \psi
-\frac{2}{9}  \bar\psi\,  [\Delta_{(1/2)} - 3/4] \, \xi
+ \frac{2}{9}  \bar\psi \, ( \hat D - 3\hat{m})\,  \psi  \right) \ .
\end{eqnarray}
Furthermore, the change in the measure is equal to  \cite{Fradkin:1983nw}
\begin{equation}
 {\cal D} \phi_\mu = {\cal D} \varphi_\mu^\bot \, {\cal D} \xi \, {\cal D} \psi\, \,
  {\rm det}^{-2}  \left[  \Delta_{(1/2)} -3/4 \right]  \ , 
\end{equation}
where the power of $-2$ comes from the fact that we are dealing with a 
two-component Dirac fermion. It follows from 
(\ref{gauge-inv}) that the components transform under a gauge transformation as 
\begin{equation}
\delta \varphi^\bot_\mu = 0\ , \quad \delta \xi = \epsilon\ , 
\quad \delta \psi = (\hat D -3\hat m)\epsilon \ .
\end{equation}
In particular, we can therefore fix the gauge $\psi=0$, for which the corresponding
Fadeev-Popov determinant is  
\begin{equation}
\label{ghost}
 \Delta_{{\rm FP}} = {\rm det}^{-2} ( \hat D -3 \hat m) \ .
\end{equation}
To  perform the one loop integration we also need to add a gauge fixing term
in the action in (\ref{action}).  This is done by  treating  
 $\hat m$ as an independent variable not given by the relation 
(\ref{gravi-mass}) in the intermediate steps of the one loop integration;
this  amounts  to adding an explicit gauge fixing term \cite{Fradkin:1983nw}.
After performing the integration over $\varphi^\bot_\mu$, $\xi$, and $\psi$ we then obtain the 
one loop determinant
\begin{eqnarray}
 Z_{{\rm 1-loop}}^{{\rm Dirac}} &=& 
{\rm det}^{-2} 
\left[  \Delta_{(1/2)} - 3/4  \right] \, {\rm det}^{-2} ( \hat D -3 \hat m)\,
   \\ \nonumber
 &&  \times {\rm det}^{2} ( \hat D -\hat m)_{\varphi^\bot} \, 
 {\rm det}^{2} (\hat D -3 \hat m)_\xi \,
 \,  {\rm det}^{2} \left[  \Delta_{(1/2)} -3/4 \right]_\xi  \,
 {\rm det}^{-2} (\hat D -3 \hat m)_\psi \ ,
\end{eqnarray}
where the first line arise from the change in the measure and the 
Fadeev-Popov determinant, while the terms in the second line come from 
integrating out $\varphi^\bot$, $\xi$ and $\psi$, as indicated by the suffices.
Simplifying and taking the square of the operators in the determinants then leads to 
(see eq.~(\ref{D.19}) and (\ref{D.20}))
\begin{eqnarray}
 Z_{{\rm 1-loop}}^{{\rm Dirac}} & = &  \frac{ {\rm det}^{2} ( \hat D -\hat m)_{(3/2)} }
{ {\rm det}^{2} (\hat D -3 \hat m)_{(1/2)} } 
\label{squaring}
= \frac{ {\rm det}( -\Delta_{(3/2)}  - \frac{9}{4} )}
{{\rm det} (  -\Delta_{(1/2)} + \frac{3}{4} )} \ .
\end{eqnarray}

The actual one loop determinant for the Majorana gravitino that appears in 
${\cal N}=1$ supergravity is  the square root of (\ref{squaring}), {\it i.e.}\ 
\begin{equation}
\label{final-gravitino}
 Z_{{\rm 1-loop }}^{\rm gravitino}
= \left( \frac{ {\rm det}( -\Delta_{(3/2)}  - \frac{9}{4} )}
{{\rm det} (  - \Delta_{(1/2)}  + \frac{3}{4} )} \right)^{1/2} \ ,
\end{equation}
and its logarithm is hence given by 
\begin{eqnarray}
\log   Z_{{\rm 1-loop}}^{{\rm gravitino}}  &=  & 
 \frac{1}{2} \log( {\rm det} ( -\Delta_{(3/2)}  -9/4) )  
- \frac{1}{2} \log ( {\rm det} ( -\Delta_{(1/2)}  +3/4))  \\ \nonumber
&=&
- \frac{1}{2}\,  \int_0^\infty \frac{dt}{t} \left( \hat{K}^{(3/2)}(\tau, \bar\tau; t) \, e^{\frac{9}{4} t}
-  K^{(1/2)}(\tau, \bar\tau; t)  \, e^{-\frac{3}{4} t} \right) \ .
\end{eqnarray}
Since we are dealing with fermions of spin $s=\frac{1}{2}$ and $s=\frac{3}{2}$, 
the heat kernels $K^{(1/2)}(\tau, \bar\tau; t)$ and $K^{(3/2)}(\tau, \bar\tau; t)$
that appear here differ slightly from (\ref{thermalh3-ker}). Indeed, for the thermal
partition function one has to impose antiperiodic boundary conditions for the fermions
along the thermal circle. In our heat kernel calculation we have summed over the images
(labelled by $m$) that describe the contribution from wrapping the thermal circle $m$ times.
Thus for fermions we need to introduce an additional factor of $(-1)^m$. With this modification,
and after performing the $t$-integral with the help of  (\ref{integral}) we then obtain 
\begin{eqnarray}
\label{1loopgravinodet}
  \log Z_{{\rm 1-loop}}^{{\rm gravitino }} &=& -  
\frac{1}{2} \sum_{m=1}^\infty \frac{(-1)^m}{m |\sin\frac{m\tau}{2}|^2}
\left[ {\textstyle \cos\left( \frac{3}{2} m\tau_1\right)  e^{-\frac{m\tau_2}{2} } 
- \cos\left( \frac{m}{2} \tau_1\right) e^{-\frac{3m\tau_2}{2}}} \right] \nonumber \\
&=&-\sum_{m=1}^{\infty}{(-1)^m\over m}\left[ {q^{3m\over 2}\over 1-q^m}
+{\bar{q}^{3m\over 2}\over 1-\bar{q}^{m}} \right] 
= \sum_{n=1}^{\infty} \log |1 + q^{n+\frac{1}{2}} |^2 \ ,
\end{eqnarray}
where the sum over $n$ comes again from the geometric series.
Thus the partition function of the ${\cal N}=1$ gravitino is given by
\begin{equation}
\label{fin-gravitino-part}
 Z_{{\rm 1-loop}}^{{\rm gravitino } }= 
\prod_{n=1}^\infty |1+ q^{n+\frac{1}{2}} |^2 \ .
\end{equation}
Together with (\ref{graviton-final}) and the tree level contribution this then gives 
\begin{equation}\label{totans}
Z_{{\rm combined} } = |q|^{-2k} 
\prod_{n=2}^\infty \frac{ | 1 + q^{n - \frac{1}{2}} |^2  }{|1- q^n|^2} \ , 
\end{equation}
where the factor $|q|^{-2k}$ is the contribution of the tree level partition function. This partition 
function has indeed the form of a trace 
\begin{equation}\label{tr}
 Z = {\rm Tr} \Bigl(  q^{L_0-\frac{c}{24}} \bar q^{\bar L_0 - \frac{\bar{c}}{24}}  \Bigr)
\end{equation}
over the  irreducible vacuum representation of the ${\cal N}=1$ super Virasoro algebra
at $c=\bar{c}=24 k$, as argued  on the basis of a quantum Brown-Henneaux reasoning in 
\cite{Maloney:2007ud}.  
\smallskip

Incidentally, if we impose instead periodic boundary conditions for the fermions along the 
thermal circle, we would obtain (\ref{1loopgravinodet}) without the factor of $(-1)^m$. 
Performing the same steps as above this would then lead to 
\begin{equation}
Z'_{{\rm combined} } = |q|^{-2k} 
\prod_{n=2}^\infty \frac{ | 1 - q^{n - \frac{1}{2}} |^2  }{|1- q^n|^2}  = 
{\rm Tr} \Bigl(  (-1)^F q^{L_0-\frac{c}{24}} \bar q^{\bar L_0 - \frac{\bar{c}}{24}}  \Bigr) \ ,
\end{equation}
which corresponds, as expected, to the introduction of a $(-1)^F$ factor in the
dual conformal field theory partition function.

\section{Final Remarks}

We have seen how the heat kernel (and therefore the one loop determinants) 
for arbitrary spin $s$ fields on (thermal) AdS$_3$ can be obtained in a group theoretic way.
The simplicity of the final answer 
\eq{thermalh3-ker}, expressed 
in terms of characters of $SL(2, {\mathbb C})$ (see \eq{h3orbheat}), 
is a reflection of the underlying symmetry of the spacetime. It is
interesting to observe that the computation of the one loop (super)gravity answers of Sec.~7 
essentially assembles  these $SL(2, {\mathbb C})$ characters into a (super) Virasoro character, 
where the  $SL(2, {\mathbb C})$ is the global part of the asymptotic isometry group given by the 
two copies of the Virasoro algebra. We therefore believe there is useful insight to be gained by 
viewing the one loop heat kernel answers in this group theoretic way.
 
Amongst the potential applications of the results given here are checks of 
the conjectures made in  \cite{Maloney:2009ck} for the one loop behaviour of 
chiral or log gravity. An explicit calculation of the one loop fluctuations of 
the chiral (log) gravity action should be amenable to a similar analysis. 

Moving further onto more nontrivial theories of gravity, 
the heat kernel can be expected to play a useful role 
in a better understanding of one loop string theory on 
AdS$_3$ \cite{Maldacena:2000kv}. This was, in fact, one 
of the prime motivations for this work. One expects the one loop string 
computation to be assembled as a sum of heat kernel contributions of 
different spin (and mass). The exact answer of
\cite{Maldacena:2000kv} does actually reflect this property. These and 
related matters are currently under investigation \cite{uswip}, and we hope 
to report on them soon.

Finally, the considerations of this paper can be generalized, 
using a similar group theoretic approach, to higher dimensional AdS
spacetimes (and their quotients). Once again, this is likely to be useful 
in the investigation of the one loop quantum string/M dynamics on these 
spacetimes. Another case of interest is AdS$_2$ where the methods of this 
paper could be useful in evaluating Sen's quantum entropy function (see, 
for instance, \cite{Sen:2008yk, Sen:2008vm, Banerjee:2009af}).

\bigskip

{\bf Acknowledgements:} We would like to thank B.~Ananthanarayan,
A.~Maloney, S.~Minwalla, J.~Teschner  and  S.~Wadia for useful 
conversations. We would also like to thank the 
International Centre for Theoretical Sciences (ICTS) for the hospitality during the 
Monsoon Workshop on String Theory at TIFR. R.G acknowledges the hospitality of 
the CTS (ETH Zurich), KITP (Santa Barbara), MPI for Gravitation, (Potsdam) and 
LPTHE (Paris) in the course of this project. The work of M.R.G.\ is partially
supported by the Swiss National Science Foundation. 
R.G.'s work was partly supported by a 
Swarnajayanthi Fellowship of the Dept.\ of Science and Technology, Govt.\ of India 
and as always by the support for basic science by the Indian people.

\section*{Appendix} 
\appendix

\section{Change of Sections as Change of Basis}

In this section we show that the tensor harmonics are completely independent of the choice 
of the section.  A different choice of section just results in a different choice of the basis
in which the tensor harmonics are expressed. We demonstrate this  by evaluating the 
tensor harmonics given in
(\ref{psidef}) for the section $\hat\sigma$, where $\hat\sigma$ is defined via 
(\ref{hsig}). Instead of (\ref{psidef}) we obtain  
\begin{eqnarray}
\hat\Psi_{a}^{(n;m_1, m_2)} (g) & = & \sum_{p_1,p_2} 
\langle s,a| \frac{n}{2} + s ,p_1;\frac{n}{2},p_2\rangle \, 
D^{(\frac{n}{2} + s )}_{p_1,m_1}(h^{-1} \cdot g_L^{-1}) \, 
D^{(\frac{n}{2} )}_{p_2,m_2}(h^{-1}\cdot g_R^{-1})
\nonumber \\
& = & \sum_{p_1,p_2} \langle s,a| \frac{n}{2}+s,p_1;\frac{n}{2},p_2\rangle \, \nonumber \\
& & \qquad \times 
\sum_{q_1,q_2} 
D^{(\frac{n}{2} +s )}_{p_1,q_1}(h^{-1})\, D^{(\frac{n}{2} )}_{p_2,q_2}(h^{-1}) \, 
D^{(\frac{n}{2}+s  )}_{q_1,m_1}(g_L^{-1}) \, D^{(\frac{n}{2} )}_{q_2,m_2}(g_R^{-1}) \ . \nonumber
\end{eqnarray}
Next we observe that 
\be{d1}
 \sum_{p_1,p_2} \langle s,a| \frac{n}{2}+s,p_1;\frac{n}{2},p_2\rangle \, 
D^{(\frac{n}{2} +s )}_{p_1,q_1}(h^{-1})\, D^{(\frac{n}{2} )}_{p_2,q_2}(h^{-1}) 
=\sum_b D^{(s)}_{ab}(h^{-1}) \langle s,b| \frac{n}{2} +s,q_1;\frac{n}{2},q_2\rangle 
\ee
since the Clebsch-Gordon coefficients describe the decomposition of the
tensor product into the spin $s$ representation. Thus we obtain
\bea{A2}
\hat\Psi_{a}^{(n; m_1, m_2)} (g) & = &
\sum_b D^{(s)}_{ab}(h^{-1})
\sum_{q_1,q_2}  \langle s,b| \frac{n}{2}+s,q_1; \frac{n}{2},q_2\rangle \, 
D^{(\frac{n}{2} )}_{q_1,m_1}(g_L^{-1}) \, D^{(\frac{n}{2} )}_{q_2,m_2}(g_R^{-1})  \nonumber \\
& = & 
\sum_b D^{(s)}_{ab}(h^{-1}) \, \Psi_{b}^{(n;m_1, m_2)} (g) \ . 
\eea
On the other hand, the basis (\ref{sigmaten}) with respect to which this
tensor harmonic is defined also changes as we change the section. In fact, it 
follows directly from  (\ref{sigmaten})  that 
\be{d2}
\hat{\bf \theta}_a(x) =  \sum_b \sigma(x) D^{(s)}_{ab}(h)\, {\bf v}_b 
=  \sum_b  D^{(s)}_{ab}(h) \, {\bf \theta}_b(x) \ .
\ee
This basis thus transforms precisely in the opposite way to the tensor harmonics,
so that 
\be{d3}
\sum_a \hat\Psi_a \, \hat{\bf \theta}_a = \sum_a \Psi_a \, {\bf \theta}_a \ .
\ee
Thus the actual tensor harmonic is completely independent of the choice
of the section, as had to be the case.

\section{Vielbeins for the Thermal  Section}

In this section, we will obtain the vielbein for the thermal section using the two 
different coordinates (\ref{spheref1}) and (\ref{torusf}). For the case
of $G=SU(2)$, a natural basis for the tangent space at the
identity of $SU(2)\times SU(2)/SU(2)$ is given by ${\bf T}_a = (T_a,-T_a)$, $a=1,2,3$, 
where
\be{basis}
T_1 = i \left( \matrix{ 0 &  i \cr -i & 0  }\right) \ , \qquad
T_2 = i \left( \matrix{ 0 &  1 \cr 1 & 0  }\right) \ , \qquad
T_3 = i \left( \matrix{ 1 &  0 \cr 0 & -1 }\right) \ .
\ee
In the coordinates (\ref{spheref1}) the thermal section is given by 
(\ref{geone1}) and (\ref{getwo1}). 
The tangent vector $\sigma(g) (T_a,-T_a)$ describes the variation
\begin{eqnarray}
g_L &\mapsto &  \tilde{g}_L  = g_L + \epsilon g_L\, T_a \\
g_R & \mapsto & \tilde{g}_R = g_R - \epsilon g_R\, T_a 
\end{eqnarray}
and this leads to 
\begin{eqnarray}
\tilde{g}_L \, \tilde{g}_R^{-1} & = & (g_L + \epsilon g_L\, T_a)\cdot 
(g_R^{-1} +\epsilon T_a  g_R^{-1})  \nonumber \\
& = & g_L \cdot g_R^{-1} + \epsilon g_L \,T_a\, g_R^{-1} + \epsilon g_L \, T_a \, g_R^{-1} 
+ {\cal O}(\epsilon^2) \ .
\end{eqnarray}
Hence the corresponding tangent vector for $(G\times G)/G$ is simply
\be{tangenta}
\delta g = g_L\, T_a \, g_R^{-1} \ .
\ee
For the above section one then finds
\begin{eqnarray}
g_L(\chi,\theta,\phi) \, T_3 \, g_R(\chi,\theta,\phi)^{-1} & = &   i\left(
\begin{array}{cc}
e^{i\chi} \cos^2 \frac{\theta}{2} - e^{-i\chi} \sin^2\frac{\theta}{2} \quad & 
 \cos\chi\, \sin\theta\, e^{i\phi} \\
\cos\chi\, \sin\theta\, e^{-i\phi}  \quad & 
e^{i\chi} \sin^2 \frac{\theta}{2} - e^{-i\chi} \cos^2\frac{\theta}{2} 
\end{array} \right) \nonumber \\
&  =  & \partial_\chi \, g  \label{T3}
\end{eqnarray}
as well as 
\be{T2}
g_L (\chi,\theta,\phi) \, T_2 \, g_R(\chi,\theta,\phi) ^{-1} = i\, \left(
\begin{array}{cc}
-\sin\theta  \quad & \cos\theta\, e^{i\phi} \\
 \cos\theta\, e^{-i\phi}  \quad & \sin\theta 
 \end{array} \right)  = \frac{1}{\sin\chi} \, \partial_\theta \, g \ ,
\ee
and
\be{T1}
g_L(\chi,\theta,\phi)  \, T_1 \, g_R(\chi,\theta,\phi) ^{-1} =\,  \left(
\begin{array}{cc}
0 & -e^{i\phi} \\
e^{-i\phi} & 0 
\end{array} \right) = \frac{1}{\sin\chi\, \sin\theta}\,  \partial_\phi \, g \ .
\ee
Thus the corresponding vielbein is the standard vielbein defined by
\be{vielc}
{\bf e}_3 = \partial_\chi \ , \qquad
{\bf e}_2 = \frac{1}{\sin\chi} \partial_\theta \ , \qquad
{\bf e}_1 = \frac{1}{ \sin\chi\, \sin\theta} \partial_\phi \ .
\ee
\smallskip

In the double polar coordinates (\ref{torusf}) the thermal section is given by 
(\ref{gLtrot}) and (\ref{gRtrot}). The same arguments as above then imply that
the corresponding vielbein is 
\be{T1t}
g_L (\psi,\eta,\varphi)\, T_1 \, g_R(\psi,\eta,\varphi)^{-1} =  \left(
\begin{array}{cc}
0 & - e^{i\varphi} \\
e^{-i\varphi} & 0 
\end{array} \right) = \frac{1}{\sin\psi}\,  \partial_\varphi\,  g \ ,
\ee
\be{T2t}
g_L(\psi,\eta,\varphi) \, T_2 \, g_R(\psi,\eta,\varphi)^{-1} =  \left(
\begin{array}{cc}
- e^{-i\eta}   \sin\psi  \quad& i e^{i\varphi}   \cos\psi  \\
i e^{-i\varphi}  \cos\psi \quad &  -e^{i\eta}  \sin\psi
\end{array} \right) = \partial_\psi\, g \ ,
\ee
and 
\be{T3t}
g(\psi,\eta,\varphi)_L \, T_3 \, g_R(\psi,\eta,\varphi)^{-1} =  i \left(
\begin{array}{cc}
e^{-i\eta}   & 0  \\
0 &  e^{i\eta}  
\end{array} \right) = - \frac{1}{\cos\psi}\, \partial_\eta\,  g \ ,
\ee
leading to 
\be{torusviel}
{\bf e}_1 =  \frac{1}{\sin\psi}\,  \partial_\varphi\ , \qquad
{\bf e}_2 = \partial_\psi \ , \qquad
{\bf e}_3 = - \frac{1}{\cos\psi}\, \partial_\eta \ .
\ee

\section{Evaluation of the Radial Heat Kernel on $S^3$}

To evaluate (\ref{heat22}) it is convenient to write $l_1=\frac{\hat{n}}{2}\pm \frac{s}{2}$, 
$l_2=\frac{\hat{n}}{2}\mp \frac{s}{2}$, where $\hn=n+s$. 
Then the Racah formula for the 
Clebsch-Gordan coefficent appearing in \eq{heat22} is particularly simple
\begin{eqnarray}
& & | \langle {\hn-s\over 2}, k; {\hn+s\over 2}, -k+a | s ,  a \rangle |^2 
= | \langle {\hn+s\over 2}, - k+a; {\hn-s\over 2},k | s ,  a \rangle |^2 \nonumber \\
& & \qquad \qquad
= \left[{(\hn-s)! (2s+1)!\over (\hn+s+1)!}  \right]
\times  {({\hn+s\over 2}-k+a)!({\hn+s\over 2}+k-a)!\over ({\hn-s\over 2}-k)!
({\hn-s\over 2}+k)! (s+a)!(s-a)!}  \ . \label{clebgordsval}
\end{eqnarray}
The sum we need to carry out --- we are suppressing for the moment the 
$k$-inde\-pen\-dent bracket $[\cdot ]$ in (\ref{clebgordsval}), as well as 
$a_{\hn}^{(s)} e^{E_{\hn}^{(s)}t}$ --- is
\begin{eqnarray}
K^{(s)}_{a;n}(\chi)& = &  {1\over (s+a)!(s-a)!}
\sum_{k=-{\hn-s\over 2}}^{\hn-s\over 2}
{({\hn+s\over 2}-k+a)!({\hn+s\over 2}+k-a)!\over ({\hn-s\over 2}-k)!({\hn-s\over 2}+k)!}  \nonumber \\
& & \qquad \qquad \qquad \qquad \qquad \qquad \times 
\left(e^{i(2k-a)\chi} + e^{-i(2k-a)\chi} \right)\ , \label{spinssumk}
\end{eqnarray}
where the two terms in the last line come from the two different choices
$l_1=\frac{\hn}{2}\pm \frac{s}{2}$ and $l_2=\frac{\hn}{2}\mp \frac{s}{2}$.(We are 
assuming here that $s>0$ --- for $s=0$ the second term is not present.)
Note that this expression is symmetric under $a\mapsto -a$, since this can be 
absorbed into relabelling $k\mapsto -k$. We may therefore, without loss of generality, 
restrict ourselves to $a\geq 0$.

Putting $p=k+{\hn-s\over 2}$, the first exponential in (\ref{spinssumk}) becomes 
\be{spinssumk2}
z^{-a}\sum_{p=0}^{\hn-s}{(p+s-a)!\over p!(s-a)!}{(\hn-p+a)!\over (\hn-s-p)!(s+a)!}z^{(2p-\hn+s)} \ ,
\ee
where we have written $z=e^{i\chi}$. To evaluate this sum let us define the generating
function
\be{sumgenfn}
F_{s,a}(w,z)= \Bigl[\sum_{p=0}^{\infty}{(p+s-a)!\over p!(s-a)!}(wz)^p\Bigr] \times
\Bigl[\sum_{q=0}^{\infty}{(q+s+a)!\over q!(s+a)!}(wz^{-1})^q \Bigr] \ ,
\ee
whose $w^{\hn-s}$ coefficient is precisely the sum in \eq{spinssumk2} (without the prefactor
of $z^{-a}$). The sums in 
(\ref{sumgenfn}) can be worked out straightforwardly, and we obtain
\bea{sumgenfn1}
F_{s,a}(w,z) &=& {1\over (1-wz)^{s-a+1}}{1\over (1-wz^{-1})^{s+a+1}}= 
{1\over [(1-wz)(1-wz^{-1})]^{s+a+1}}(1-wz)^{2a} \cr
&=& {1\over(1-2w\cos{\chi}+w^2)^{s+a+1}}(1-wz)^{2a} \ .
\eea
The first term in $F_{s,a}(w,z)$ is precisely the generating function for the 
Gegenbauer polynomials 
\be{gegengen}
{1\over(1-2w\cos{\chi}+w^2)^{\lambda}}=\sum_{p=0}^{\infty}C_p^{\lambda}(\cos{\chi})\, w^p
\ee
(see 8.930 of \cite{GR}), and thus we find for (\ref{spinssumk}) 
\be{spinssumk3}
K^{(s)}_{a;n}(\chi)=(2-\delta_{s,0})\,  \sum_{r=0}^{min(2a,n)}(-1)^r{(2a)!\over r! (2a-r)!}
\cos[(r-a)\chi] \, C_{n-r}^{s+a+1}(\cos{\chi}) \ ,
\ee
where we have now restored the $z^{-a}$ term from (\ref{spinssumk2}) and 
included the second exponential in (\ref{spinssumk}),
{\it i.e.}\ added in the term with $\chi\mapsto -\chi$. (For prefactor 
$(2-\delta_{s,0})$ guarantees that the result is also correct for $s=0$.) 
 In addition we have used that 
$\hn -s = n$. We note in passing that 
for $a=0$ this simplifies to $K^{(s)}_{0;n}(\chi) = (2-\delta_{s,0}) \, C_{n}^{s+1}(\cos{\chi})$.
We also remind the reader that this expression is only valid for $a\geq 0$, and
that $K^{(s)}_{a;n}(\chi)$ is invariant under $a\mapsto -a$. 

Including the prefactors that were left out in going to (\ref{spinssumk}) we then 
obtain for (\ref{heat22})
\be{spinsheatkdiagfin}
K^{(s)}_{a}(\chi,t)= \frac{1}{2 \pi^2}\, 
\sum_{n=0}^{\infty} 
{(n+1)! (2s)!\over (n+2s)!}  \, K^{(s)}_{a;n}(\chi)\, e^{-((n+s)(n+s+2)-s )t}\ .
\ee
In the scalar case, $s=0$, we have $a=0$, and the formula agrees with
(\ref{scalk}) since the first Gegenbauer polynomial simply equals
\be{Geg}
C^1_{n}(\cos\chi) = \frac{\sin(n+1)\chi}{\sin\chi} \ .
\ee

\section{Gravitino Action}

In this appendix we provide the details for the derivation of the action (\ref{action}). 
We start with the gravitino Lagrangian (\ref{lag}), and express $\psi$ in terms of 
$\varphi^\bot$, $\xi$, and $\psi$, using (\ref{ex1}) and (\ref{expansion}). The resulting
terms are all quadratic in these fields, and we shall analyze them in turn.
\smallskip

\noindent \underline{The quadratic term in $\varphi^\bot$} is given by 
\begin{equation}
\label{lag1}
 - \int d^3z \sqrt{g} \, \bar\varphi^\bot _\mu  (  \Gamma^{\mu\nu\rho} D_\nu   
+ \hat m \Gamma^{\mu\rho} )
\varphi^\bot_\rho \ ,
\end{equation}
where in Euclidean space $\bar\varphi^\bot = ( \varphi^{\bot})^\dagger$. Using
that $\Gamma^\mu \varphi^\bot_\mu=0$ as well as  
$\{\Gamma^\mu, \Gamma^\nu\} = 2 g^{\mu\nu}$ and the definition (\ref{7.14}),
we find
\begin{equation}
\label{term-1}
-\int d^3 z \sqrt{g}\,  \bar\varphi^{\bot\mu}(  \Gamma^\nu D_\nu  - \hat m ) \varphi^\bot_\mu \ .
\end{equation}
\smallskip

\noindent \underline{The cross term between $\varphi^\bot$ and $\xi$} is of the form
\begin{equation}
\label{left-over}
- \int d^3 z \sqrt{g} \,   \left(\bar\varphi^{\bot\rho} \Gamma^\mu D_\mu D_\rho \, \xi
+ D^\rho \xi^\dagger \Gamma^\mu D_\mu \varphi^\bot_\rho \right) \ ,
\end{equation}
where we have used that $\Gamma^\mu \varphi^\bot_\mu =D^\mu \varphi^\bot_\mu =0$. 
Both terms actually vanish. For the first term we use
\begin{equation}
\label{spinhalf}
  (D_{\mu} D_{\rho} - D_{\rho} D_{\mu} ) \xi = 
\frac{1}{8} R_{\mu\rho \sigma\delta} [\Gamma^\sigma, \Gamma^\delta]
\xi
\end{equation}
to move $D_\rho$ to the left of $D_\mu$, where it vanishes (up to a total derivative) since 
$D_\rho \varphi^{\bot\rho}=0$. Thus the first term equals 
\begin{equation}
 -\frac{1}{8} \int d^3 z \sqrt{g}\,  \bar\varphi^{\bot \rho} \Gamma^\mu R_{\mu\rho\sigma\delta} 
[\Gamma^\sigma,  \Gamma^\delta] \xi \ ,
\end{equation}
which is seen to vanish upon using (\ref{curv}) and 
$\Gamma^\mu \varphi^\bot_\mu =0$. Similar manipulations can be used to show that the 
second term in  (\ref{left-over}) also vanishes.
\smallskip

\noindent \underline{The cross term between $\varphi^\bot$ and $\psi$} vanishes
directly upon using $D^\mu\varphi^{\bot}_\mu=\Gamma^\mu \varphi^\bot_\mu =0$.
\smallskip

\noindent \underline{The quadratic term involving the spinor component $\xi$} arises from
\begin{equation}\label{D.6}
 - \int d^3 z \sqrt{g}\, (\overline{\tilde D_\mu \xi}) \, (  \Gamma^{\mu\nu\rho }D_\nu + \hat m 
\Gamma^{\mu\rho} ) \tilde{D}_\rho  \xi \ ,
\end{equation}
where $\tilde{D}_\rho = D_\rho - \frac{\Gamma_\rho}{3} \hat{D}$ is the differential operator
that appeared in the defining equation for $\xi$, (\ref{expansion}). Using 
$\Gamma^\mu \tilde D_\mu \xi =0$, and performing the same steps as in the analysis leading
to (\ref{term-1}), we can rewrite (\ref{D.6}) as 
\begin{equation}
\label{xiaction}
 - \int d^3 z \sqrt{g} \, (\overline{ \tilde D^\mu \xi} )\, ( \Gamma^\rho D_\rho - \hat m ) \tilde D_\mu \xi \ .
\end{equation}
Next we integrate by parts to move the operator $\tilde D^\mu$ to the right. Using 
$\Gamma^\mu\tilde{D}_\mu \xi=0$ the term proportional to $\hat m$ reduces to 
\begin{equation}
  -  \hat m\int d^3 z\sqrt{g}\left(  \bar\xi \, D_\mu  
  (  D^\mu  - \frac{\Gamma^\mu}{3} \hat D ) \xi \right) \ ,
\end{equation}
where we have written out $\tilde{D}^\mu$ in terms of the covariant derivative $D^\mu$ and
$\hat{D}$. For the first term in (\ref{xiaction}) integration by parts leads to 
\begin{equation}
 \underbrace{
 \int d^3 z \sqrt{g} \, \bar\xi \, D_\mu ( \Gamma^\sigma D_\sigma) 
 \tilde{D}^\mu \xi}_{A}\;
\underbrace{- \frac{1}{3}\int d^3 z \sqrt{g}\, \bar\xi 
 (\Gamma^\rho   D_\rho) \Gamma_\mu( \Gamma^\sigma D_\sigma) \tilde{D}^\mu 
\xi}_{B}   \ .
\end{equation}
For $B$ we use $\{\Gamma_\mu , \Gamma^\sigma \} = 2 \delta_\mu^\sigma$
as well as $\Gamma_\mu\tilde{D}^\mu \xi=0$ to obtain
\begin{equation}\label{D.10}
 B = - \frac{2}{3} \int d^3 z \sqrt{g} \, \bar\xi (\Gamma^\rho  D_\rho) D_\mu 
 \left( D^\mu - \frac{1}{3} \Gamma^\mu \hat D\right) \xi \ .
\end{equation}
For $A$ we use the commutation relation
\begin{equation}
( D_\mu D_\sigma - D_\sigma D_\mu) \tilde D^\mu \xi 
= R_{\mu \sigma}  \tilde D^\mu \xi  + \frac{1}{8} R_{\mu\sigma \nu\delta} 
[\Gamma^\nu, \Gamma^\delta]  \tilde D^\mu \xi \ .
\end{equation}
to rewrite it as 
\begin{equation}\label{D.12}
 A 
= \int d^3 z \sqrt{g} \, \left(  \bar\xi  ( \Gamma^\sigma D_\sigma) D_\mu
\tilde{D}^\mu \xi + 
\bar\xi \,   \Gamma^\sigma R_{\mu \sigma}  \tilde D^\mu \xi 
+ \bar\xi  \, 
 \Gamma^\sigma  \frac{1}{8} R_{\mu\sigma\nu\delta} 
 [\Gamma^\nu, \Gamma^\delta]  \tilde D^\mu  \xi \right)  \ .
\end{equation}
Substituting the explicit expressions (\ref{curv}) for the 
the curvature tensor and Ricci tensor of $H_3^+$, the last two terms of (\ref{D.12})
become
\begin{equation}
\frac{R}{3} \int d^3 z \sqrt{g}\,  \left( \bar\xi\, \Gamma^\mu \tilde D_\mu \xi
-\frac{1}{2} \bar\xi\, \Gamma^\mu \tilde D_\mu \xi \right)= 0 \ , 
\end{equation}
which vanish because of $\Gamma^\mu \tilde D_\mu \xi =0$. The first term
of $A$ in (\ref{D.12}) has the same form as $B$ in (\ref{D.10}), and thus 
the total contribution quadratic in $\xi$ equals
\begin{equation}
\int d^3 z \sqrt{g} \,     \left[
 \frac{1}{3} \bar\xi\, 
 ( \Gamma^\sigma D_\sigma) D_\mu \left(  D^\mu - \frac{\Gamma^\mu}{3} \hat D \right)  \xi
  -  \hat m  \, \bar\xi  D_\mu   \left(  D^\mu  - \frac{\Gamma^\mu}{3} \hat D  \right) \xi \right]\ .
\end{equation}
Using (\ref{spinhalf}) we can simplify 
\begin{equation}
\label{def-deltahalf}
D_\mu  \left( D^\mu  - \frac{\Gamma^\mu}{3} \hat{D} \right) \xi = 
D_\mu \left( D^\mu - \frac{1}{3} \Gamma^\mu \Gamma^\sigma D_\sigma \right) \xi
= \frac{2}{3} \left( \Delta_{(1/2)} + \frac{R}{8} \right) \xi \ .
\end{equation}
Thus the final answer for the quadratic $\xi$ term takes the form
\begin{eqnarray}
\label{term-2}
 \frac{2}{9} \int d^3z \sqrt{g}\, \bar\xi ( \Gamma^\sigma D_\sigma - 3\hat m)
(\Delta_{(1/2)} +R/8) \xi \ .
\end{eqnarray}
\smallskip

\noindent \underline{The cross term between $\xi$ and $\psi$} can be analyzed similarly, 
and it leads to 
\begin{equation}
\label{term-3}
 \frac{2}{9} \int d^3 z \sqrt{g} \, 
 \Bigl[  \bar\psi ( \Delta_{(1/2)} + R/8)  \xi - \bar\xi ( \Delta_{(1/2)} + R/8) \psi \Bigr]    \ .
\end{equation}
\smallskip

\noindent \underline{The quadratic term in $\psi$}  reduces with similar manipulations to 
\begin{equation}
\label{term-4}
 - \frac{2}{9} \int d^3 z \sqrt{g} \, \bar\psi ( \hat D + 3 \hat m ) \psi \ .
\end{equation}
Combing (\ref{term-1}), (\ref{term-2}), (\ref{term-3})  and (\ref{term-4}), and setting $R=-6$ then
finally leads to eq.~(\ref{action}). 
\medskip

\noindent \underline{For the derivation of (\ref{squaring})} we also need the identities
 \begin{equation}\label{D.19}
- (\Gamma^\mu D_\mu +\hat m ) ( \Gamma^\rho D_\rho - \hat m)\, \varphi^\bot_\sigma
= {\textstyle ( - D^\mu D_\mu + \frac{5R}{12} + \hat m^2)\, \varphi^\bot_\sigma 
=  (-\Delta_{(3/2)} - \frac{9}{4}) \, \varphi^\bot_\sigma }
\end{equation}
and
\begin{equation} 
\label{D.20}
- ( \Gamma^\sigma D_\sigma + 3\hat m)( \Gamma^\rho D_\rho -3\hat m)\, \xi
= {\textstyle ( -  D^\mu D_\mu  + \frac{R}{4} + 9 \, \hat m^2)\, \xi 
=  (-\Delta_{(1/2)} + \frac{3}{4}) \, \xi }\ . 
\end{equation}
They follow upon using (\ref{spinhalf}) and the analogue for spin $3/2$ 
\begin{equation}
\label{square32}
(D_\mu D_\rho - D_\rho D_\mu) \varphi_{\nu}^\bot  
=  R^{\sigma}_{\;\nu\rho\mu} \varphi^\bot_\sigma  
+ \frac{1}{8} R_{\mu\rho\sigma\delta}
 [\Gamma^{\sigma},  \Gamma^\delta] \varphi^\bot_\nu \ ,
\end{equation}
as well as (\ref{curv}). We have also substituted the value of $\hat m^2$  from (\ref{gravi-mass}).

\bibliographystyle{JHEP}

\end{document}